\documentclass[aps,prd,nofootinbib,tightenlines,notitlepage,floatfix,superscriptaddress,showkeys,onecolumn]{revtex4-1}

\RequirePackage{fix-cm}

\RequirePackage{color,graphicx,url}

\usepackage{placeins}

\let\Oldsection\section
\renewcommand{\section}{\FloatBarrier\Oldsection}

\let\Oldsubsection\subsection
\renewcommand{\subsection}{\FloatBarrier\Oldsubsection}

\let\Oldsubsubsection\subsubsection
\renewcommand{\subsubsection}{\FloatBarrier\Oldsubsubsection}

\usepackage{amsfonts,amsmath,amssymb,bm,xspace}
\usepackage{multirow}
\usepackage{longtable}
\usepackage{colortbl}
\usepackage{bm}% bold math
\usepackage{enumitem}%enumarting lists
\usepackage{nccmath}
\usepackage{xfrac}

\usepackage{placeins}

\newcommand{\sat}{\mathrm{sat}}
\newcommand{\sym}{\mathrm{sym}}

\newcommand{\Lcal}{\mathcal{L}}
\newcommand{\psib}{\bar{\psi}}

\graphicspath{{Figures}}

\usepackage[colorlinks=true]{hyperref}  

\newcommand{\orcid}[1]{\href{https://orcid.org/#1}{\textcolor[HTML]{A6CE39}{\aiOrcid}}}
\usepackage{xcolor}

\hypersetup{%
    pdfsubject=Paper,
    pdfkeywords={nuclear physics} {MBPT} {chiral EFT} {asymmetric nuclear matter}
    unicode=true,
    breaklinks=true,
     colorlinks   = true,
     linkcolor = blue,
     citecolor = blue,
     menucolor = blue,
     urlcolor = blue
}

\begin{document}

%\title{Nuclear structure within the relativistic mean field approach including chiral symmetry and quark confinement effects}
\title{Nuclear structure within a relativistic mean field approach including chiral symmetry and confinement-inspired nucleon response}

\author{M.~ Chamseddine}
\email{mohamad.chamseddine@ijclab.in2p3.fr}
\affiliation{IJCLab, Universit\'e Paris-Saclay, CNRS/IN2P3, 91405 Orsay Cedex, France}

\author{J.-P.~Ebran}
%\email{jean-paul.ebran@cea.fr}
\affiliation{CEA,DAM,DIF, F-91297 Arpajon, France}
\affiliation{Universit\'e Paris-Saclay, CEA, Laboratoire Mati\`ere en Conditions Extr\^emes, 91680, Bruy\`eres-le-Ch\^atel, France}

\author{E.~ Khan}
%\email{elias.khan@ijclab.in2p3.fr}
\affiliation{IJCLab, Universit\'e Paris-Saclay, CNRS/IN2P3, 91405 Orsay Cedex, France}
\affiliation{Institut Universitaire de France (IUF)}

\author{B.~K.~Pradhan}
%\email{b-k.pradhan@ip2i.in2p3.fr}
\affiliation{Institut de Physique des 2 infinis de Lyon, CNRS/IN2P3, Universit\'e de Lyon, Universit\'e Claude Bernard Lyon 1, F-69622 Villeurbanne Cedex, France}

\author{J.~Margueron}
%\email{jerome.margueron@cnrs.fr}
\affiliation{International Research Laboratory on Nuclear Physics and Astrophysics, Michigan State University and CNRS, East Lansing, MI 48824, USA}

\author{H.~ Hansen}
%\email{hansen@ipnl.in2p3.fr}
\affiliation{Institut de Physique des 2 infinis de Lyon, CNRS/IN2P3, Universit\'e de Lyon, Universit\'e Claude Bernard Lyon 1, F-69622 Villeurbanne Cedex, France}

\author{G.~ Chanfray}
%\email{g.chanfray@ipnl.in2p3.fr}
\affiliation{Institut de Physique des 2 infinis de Lyon, CNRS/IN2P3, Universit\'e de Lyon, Universit\'e Claude Bernard Lyon 1, F-69622 Villeurbanne Cedex, France}

\begin{abstract}
The relativistic mean field approach, within a theoretical framework known as the chiral confining model, that incorporates both chiral symmetry breaking and confinement-inspired nucleon response, is applied for the first time to study the properties of some spherical nuclei, such as binding energies and charge radii. The model parameters are calibrated using a Bayesian approach to reproduce nuclear empirical properties and global properties of doubly magic nuclei. 
For intermediate mass and heavy closed-shell nuclei, this model provides a satisfactory description of binding energies and charge radii, with standard deviations comparable to usual relativistic mean field models, while larger discrepancies are observed in light nuclei. This behavior is traced back to the constrained form of the chiral potential, which reduces flexibility away from saturation density and affects systems probing a broader density range, i.e., light nuclei. 
Charge radii are reproduced with excellent accuracy across all mass regions considered, although the corresponding density profiles remain slightly more diffuse than the experimental ones.
Extending the analysis to open-shell nuclei using a separable Gogny pairing interaction reveals an enhanced pairing effect in this framework, which can lead to anomalous pairings even in closed shells. This is associated with the larger Dirac and non-relativistic effective masses predicted by RMF-CC1 compared with standard RMF parametrizations such as NL3 and DD-ME2, which are respectively associated with reduced spin-orbit splittings and an increased density of single-particle states around the Fermi surface. A modest reduction of the pairing strength suppresses this anomaly at the diagnostic level, indicating that the pairing channel remains sensitive to the underlying single-particle structure generated by the model.
Finally, we qualitatively explore departures from the linear sigma model potential, motivated by the Nambu–Jona-Lasinio framework, to understand the role of the chiral potential for finite nuclei properties. Allowing for variations of the cubic and quartic terms in the potential leads to improved results for light nuclei, and a reduction of the Dirac mass, which in turn suppresses the anomalous pairing. These results confirm the sensitivity of light nuclei to the global trend of the chiral potential, and the impact of the latter on the Dirac mass and thus the single-particle structure.
\end{abstract}

\maketitle

\section{Introduction}
\label{sec:intro}

The study of nuclear systems, from finite nuclei (FN) to nuclear matter, remains one of the most challenging problems in physics. It requires state-of-the-art many-body methods together with a proper understanding of the strong interaction in the relevant density and energy regimes. The nuclear interaction emerges as the residual low-energy manifestation of the strong interaction in its non-perturbative regime. While Quantum Chromodynamics (QCD) describes interactions between quarks and gluons, nuclear structure models are commonly formulated in terms of effective interactions between nucleons.
This non-perturbative regime manifests itself through highly non-linear phenomena, such as spontaneous chiral symmetry breaking, which endows the QCD vacuum with a non-trivial structure, as well as color confinement.

Unlike QED, where the main difficulty lies in the many-body problem since the interaction is simple and perturbative over a wide range of energies, the low-energy regime relevant for nuclear physics requires the use of effective approaches. These include QCD-inspired phenomenological models at the quark level, such as the Nambu–Jona-Lasinio model (NJL) model~\cite{Nambu:1961fr}, as well as systematically constructed effective theories based on the symmetries of QCD, such as Chiral Effective Field Theory ($\chi$-EFT)~\cite{Weinberg}. For medium and heavy nuclei, where fully microscopic treatments remain computationally demanding, energy-density-functionals (EDF) approaches provide an efficient and accurate framework for global calculations. Within these EDFs, relativistic mean field (RMF) theory has been successful in describing nuclear structure. In addition, it is a covariant approach in which spin degrees of freedom, the spin-orbit interaction, and causality are naturally built-in in the formalism~\cite{RING1996,VRETENAR2005,Meng2006}. However, the Lagrangian used in the RMF approach
remains phenomenological, with a weak or inexistent connection with the low-energy QCD regime. 

In this work, we use the chiral confining model (CCM) framework, taken at the Hartree level, referred to as RMF-CC. RMF-CC is not an ab initio chiral-EFT construction; it is a QCD-motivated~\cite{Chanfray2005} effective covariant functional whose scalar sector is constrained by chiral symmetry. It additionally encodes confinement-inspired nucleon response through a scalar polarizability. Here, "confinement-inspired" refers to the response of the nucleon substructure generated by a confining potential for constituent quarks, and should not be understood as a description of the fundamental color-confinement mechanism of QCD. 

It has been previously applied to describe nuclear matter properties~\cite{Massot2008,Rahul2022,Chamseddine,Chamseddine_NJL}. This model is based on two main ideas: the identification of the nuclear physics $\sigma$ meson, introduced in relativistic mean fields approaches, with the radial fluctuation of the chiral quark condensate, as proposed in Ref.~\cite{Chanfray2001}, and the inclusion of a nucleon "polarization" effect induced by the nuclear environment, as in Refs.~\cite{Guichon1988,Chanfray2005,Chanfray2006}.

The paper is organized as follows: In Section~\ref{sec:framework}, we present the formalism framing the theoretical framework of our model. In Section~\ref{sec:parametrisation}, we discuss the parameters of our model and the Bayesian procedure followed to calibrate these parameters. We then present our results in Section~\ref{sec:results}, where we discuss doubly magic and open-shell nuclei. In Section~\ref{sec:systematics}, we perform a systematic study of the spin-orbit potential within the RMF-CC phase space, examining both the constraints imposed by nuclear empirical parameters (NEPs) and the mechanisms governing its magnitude. In Section~\ref{sec:NJL}, we explore approaches beyond the Linear Sigma Model (L$\sigma$M) and explore the consequences for finite nuclei properties. Our results are summarized, and we present our conclusions in Section~\ref{sec:conclusions}.

\section{Framework}
\label{sec:framework}

\subsection{Relativistic mean field Lagrangians}

Considering only the lightest $u$ and $d$ quarks and the flavor number $N_f=2$, the chiral fields associated with the fluctuations of the quark condensate $\langle \bar q q\rangle$ resulting from chiral symmetry breaking are usually parameterized in terms of a $\rm{SU}(2)$ matrix $M$ as:
\begin{equation}
M=\sigma + i\vec{\tau}\cdot\vec{\phi}\equiv S\, U \, ,
\label{REPRES}
\end{equation}
with $S = s + f_\pi$ and $U=e^{i\,{\vec{\tau}\cdot\vec{\pi}}/{f_\pi}}$, $\vec{\tau}$ being the Pauli matrices in isospin space. The scalar field $\sigma$ ($S$) and pseudo-scalar fields $\vec{\phi}$ ($\vec{\pi}$) written in Cartesian (polar) coordinates appear as the dynamical degrees of freedom. One may be tempted to identify the usual sigma meson of the Walecka model (let us call it $\sigma_W$ from now on) with the scalar field $\sigma$ in Cartesian coordinates. It is, however, forbidden by chiral constraints, and this point has been first addressed by Birse~\cite{Birse94}: it would lead to the presence of terms of order $m_\pi$ in the nucleon-nucleon (NN) interaction, which is not allowed as discussed in detail in \cite{Chanfray2006} (see also \cite{Chanfray-Schuck} for a more recent review).

In this study, we follow Ref.~\cite{Chanfray2001} and identify $\sigma_W$ with the chiral invariant $s$ ($=S-f_\pi$) field associated with the radial fluctuation of the chiral condensate around the chiral radius $f_\pi$, in polar coordinates. It formally consists of promoting the chiral invariant scalar field $s$ and the pion field $\vec{\pi}$ appearing in the matrix $M$ in Eq.~\eqref{REPRES}
to effective degrees of freedom. This was originally formulated in the framework of the Linear Sigma Model (L$\sigma$M)~\cite{Chanfray2001,Chanfray2005,Chanfray2007,Massot2008,Massot2009}, but an explicit construction using a bosonization technique of the chiral effective potential can be done within the NJL model~\cite{Chanfray2011} where the L$\sigma$M potential is recovered through a second order expansion in  $S^2-f^2_\pi$ of the constituent quark Dirac sea energy. The latest versions of this chiral potential from NJL are used in Refs.~\cite{Chamseddine_NJL,Chanfray-Schuck,chanfray-universe}.

This proposal, which gives a plausible answer to the long standing problem of the chiral status of Walecka theories, also has the merit of respecting all the desired chiral constraints~\cite{Birse94}. In particular, the correspondence $s\equiv \sigma_W$ generates a coupling of the scalar field to the derivatives of the pion field, as expected in the physical world. Hence, the radial mode decouples from low-energy pions whose dynamics is governed by chiral perturbation theory. A detailed discussion of this somewhat subtle topic is given in Refs. \cite{Chanfray2001,Chanfray2006,Hebeler2013}, and in the following, we employ the notation $s$ for the scalar field to avoid confusion with the meson $\sigma$.

The relativistic Lagrangian can generically be written as the sum of a kinetic fermionic term and meson-nucleon term. The kinetic term is
\begin{equation}
\label{eq:L_kinetic}
\Lcal_N = \psib \left( i \gamma^{\mu}\partial_{\mu} -M_N(s) \right) \psi \, ,
\end{equation}
where the field $\psi$ represents the nucleon spinor, and $M_N(s)$ is the polarised nucleon mass defined in \eqref{eq:nucleon_mass}. 
The mesonic degrees of freedom are the chiral-invariant scalar field $s$ and the isoscalar/isovector vector fields $\omega$ and $\rho$, together with the Coulomb field. The meson-nucleon Lagrangian density then is
\begin{align}
\label{eq:L_m}
\Lcal_m \;=\;& 
\frac{1}{2}\partial^\mu s\,\partial_\mu s - V_\chi(s) \nonumber \\
&- g_\omega \omega_\mu \bar{\psi}\gamma^\mu\psi
+ \frac{1}{2} m_\omega^2 \omega^\mu\omega_\mu
- \frac{1}{4} F^{\mu\nu}F_{\mu\nu} \nonumber \\
&- g_\rho \rho_{a\mu}\bar{\psi}\gamma^\mu\tau_a\psi
+ g_\rho \frac{\kappa_\rho}{2M_N}
\partial_\nu\rho_{a\mu}\bar{\psi}\sigma^{\mu\nu}\tau_a\psi \nonumber \\
&+ \frac{1}{2} m_\rho^2 \rho_{a\mu}\rho_a^{\mu}
- \frac{1}{4} G_a^{\mu\nu}G_{a\mu\nu} \nonumber \\
&- eA_\mu \bar \psi \gamma^\mu \tfrac12(1-\tau_3)\psi - \frac14 H^{\mu \nu}H_{\mu \nu} \nonumber \\
&+ \frac{g_A}{2f_\pi}
\partial_\mu\varphi_{\pi a}\bar{\psi}\gamma^\mu\gamma^5\tau_a\psi
- \frac{1}{2} m_\pi^2 \varphi_{\pi a}\varphi_{\pi a} \nonumber \\
& + \frac{1}{2}\partial^\mu\varphi_{\pi a}\partial_\mu\varphi_{\pi a}\, .
\end{align}
where the symbols have their usual meaning. We remind that $g_A = 1.25$ is the axial coupling constant and $f_\pi = 94$ MeV is the pion decay constant. We also note the nucleon-$\rho$ tensor coupling $f_\rho = g_\rho \kappa_\rho$. The pion and tensor $\rho$ interactions do not contribute at the Hartree level, and thus, they will be removed in the following.

The chiral potential $V_\chi(s)$ in the scalar field $s$ has a typical Mexican hat shape for spontaneous symmetry breaking, where $s$ measures the chiral symmetry breaking. We consider in this study the L$\sigma$M expression:
\begin{eqnarray}
\label{eq:LsM}
V_\chi(s)&=& \frac{\lambda}{4}\big((f_\pi+s)^2-v^2\big)^2-f_\pi m_\pi^2s\nonumber\\
&\equiv &\frac{m^{2}_s}{2}s ^{2}+
\frac{m^{2}_s-m^{2}_{\pi }}{2f_\pi}s^3+
\frac{m^{2}_s-m^{2}_{\pi }}{8f_\pi^2}s^4.
\end{eqnarray}
More details on this effective chiral potential can be found, for instance, in Ref~\cite{Rahul2022} and in references therein.

In the presence of the nuclear scalar field, nucleons get polarized and their mass is modified according to:
\begin{equation}
\label{eq:nucleon_mass}
M_N(s)=M_N+g_s s+\frac{1}{2}\kappa_\mathrm{NS}\left(s^2+\frac{s^3}{3f_\pi}\right).
\end{equation}
The nucleon polarisability parameter $\kappa_\mathrm{NS}$, incorporates the effect of the nucleon response, i.e., the central ingredient of the quark-meson coupling model (QMC) introduced in the original pioneering work of P. Guichon \cite{Guichon1988}. As in Ref.~\cite{Chanfray2005} the scalar field dependent susceptibility is expressed as:
\begin{equation}
\label{eq:kappa_tilde}
\tilde\kappa_\mathrm{NS}({\bar s})\equiv {\partial^2M_N(s)\over\partial s^2}\vert_{s={\bar s}}=
\kappa_\mathrm{NS}\left(1+{{\bar s}\over f_\pi}\right) \, ,
\end{equation} 
which vanishes at full chiral restoration, i.e., $\bar s=-f_\pi$, where $\bar s$ is the value taken by the $s$ field in the ground state. We usually work with the dimensionless parameter $C_\mathrm{NS}$ defined as,
\begin{equation}
\label{eq:C_param}
C_\mathrm{NS} \equiv \frac{\kappa_\mathrm{NS} F^2_\pi}{2M_N}.
\end{equation}

\subsection{Field equations and Hamiltonian}

In the mean field approximation, the mesons take their ground state expectation values: $s \mapsto \bar s,~ \omega^\mu \mapsto \bar \omega^0 \equiv \bar \omega,~ \rho_a^\mu \mapsto \bar \rho_3^0 \equiv \bar \rho, A^\mu \mapsto \bar A^0 \equiv \bar A$. We can now list the equations of motion of the different contributing meson fields:
\begin{eqnarray}
\label{eq:classical_EOM}
&& -\nabla^2 \bar s(\mathbf r) +V_\chi'(\bar s)=-\frac{d M_N(s)}{ds} \Big |_{\bar s} \left\langle \bar\Psi\Psi\right\rangle (\mathbf r) = -g_s^*(\bar s) \left\langle \bar\Psi\Psi\right\rangle (\mathbf r)\nonumber\\
&& -\nabla^2 \bar\omega(\mathbf r) + m^{2}_{\omega}	\bar\omega(\mathbf r)=g_\omega\left\langle \Psi^\dagger\Psi\right\rangle(\mathbf r)\nonumber\\
&& -\nabla^2 \bar\rho(\mathbf r) +m^{2}_{\rho}\,	\bar\rho(\mathbf r)=g_\rho\,\left\langle \Psi^\dagger\tau_3\Psi\right\rangle(\mathbf r), \nonumber \\
&& -\nabla^2 \bar A(\mathbf r) = e \left\langle \Psi^\dagger\frac{1-\tau_3}{2}\Psi\right\rangle(\mathbf r)
\end{eqnarray}
where we define the following quantities:
\begin{align}
\label{eq:densities}
    &\left\langle \bar\Psi\Psi\right\rangle (\mathbf r) = \rho_s(\mathbf r) = \rho_{s,p}(\mathbf r) + \rho_{s,n}(\mathbf r)~, \nonumber \\
    &\left\langle \Psi^\dagger\Psi\right\rangle(\mathbf r) = \rho_b(\mathbf r) = \rho_{b,p}(\mathbf r) + \rho_{b,n}(\mathbf r)~, \nonumber \\
    & \left\langle \Psi^\dagger\tau_3\Psi\right\rangle(\mathbf r) = \rho_{b}^{(3)} =  \rho_{b,n}(\mathbf r) - \rho_{b,p}(\mathbf r),\nonumber \\
    & \left\langle \Psi^\dagger\frac{1-\tau_3}{2}\Psi\right\rangle(\mathbf r)=\rho_{c}(\mathbf r)~,
\end{align}
where $p/n$ stands for protons/neutrons and $\tau_3(n/p) = +1/-1$ respectively.

The equation of motion for the nucleon can be written as:
\begin{align}
\label{eq:Dirac_eqn}
    \left[
-i\gamma^\mu \partial_\mu
+
\left(M_N + \Sigma_S(\mathbf{r})\right)
+
\gamma_0 \Sigma^0(\mathbf{r})
\right]
\psi(\mathbf r)
=
0~,
\end{align}
with the scalar and time-like component potentials defined as:
\begin{align}
    &\Sigma_S(\mathbf r) = M_N(\bar s) -M_N~, \\
    &\Sigma_0(\mathbf r) = g_\omega \bar \omega(\mathbf r) + g_\rho  \tau_3 \; \bar \rho(\mathbf r)+ \frac12 e(1-\tau_3)\bar A(\mathbf r)~.
\end{align}

We can finally define the hamiltonian density
\begin{align}
    \mathcal H(\mathbf r) =& \bar \Psi (-i \gamma^i \partial_i + M_N) \Psi +  \Big(M_N(\bar s) -M_N \Big)\rho_s(\mathbf r) + V_\chi(\bar s) - \tfrac12 \bar s V^{'}_\chi(\bar s) - \tfrac12 \bar s g_s^*(\bar s) \rho_s(\mathbf r) \nonumber \\
    &+ \frac12 \Big(g_\omega \bar \omega \rho_b(\mathbf r) + g_\rho \bar \rho \rho_b^{(3)}(\mathbf r) + e \bar A \rho_c(\mathbf r) \Big)~, 
\end{align}
and the total energy $E$ is obtained as: $E=\int \! d{\mathbf r} \, \mathcal H(\mathbf r)$.

\section{Parametrization and fitting procedure}
\label{sec:parametrisation}

In this study, we will employ a Bayesian approach using the Markov Chain Monte-Carlo (MCMC) method. In this way, the full exploration of the uncertainties in the experimental data is translated into uncertainties in the free parameters of the model and in other predicted quantities that we will see in the results.

In the Bayesian approach, the so-called "data" refer to experimental data, e.g., binding energies $B$, charge radii $R_c$, and pseudo-data such as the NEPs: the saturation density $n_\sat$, saturation energy $E_\sat$, symmetry energy $E_\sym$, and the incompressibility modulus $K_\sat$ (see Refs.~\cite{Margueron2018,Margueron2019} for more details). A certain "parameterization" of the model refers to the model's parameters, i.e., quantities like coupling constants, e.g., $g_\omega$, $g_s$, and $C_\mathrm{NS}$, or masses, e.g., $m_s$, which directly enter into the definition of the model. A "parameterization" can therefore be simply represented by the set $ \{ \theta_i \} $ of these parameters. They will be fixed in such a way as to reproduce the previously mentioned ``data". 
The Bayesian probability associated with a certain parameterization and a set of data can be obtained in two different ways, since (Bayes theorem):
\begin{equation}
P(\{\theta_i\} \mid \textrm{data}) \times P(\textrm{data}) =  P(\textrm{data}\mid \{\theta_i\})\times P(\{\theta_i\}),
\label{eq:bayes}
\end{equation}
where $P(\{\theta_i\} \mid \textrm{data})$ is the posterior probability associated with a parameterization given a set of ``data", $P(\textrm{data})$ is the evidence, which is defined as the normalization of the posterior probability.
We have $P(\textrm{data}\mid \{\theta_i\})$, the likelihood probability associated with a set of ``data" given a parameterization, which can be expressed, for instance, as
\begin{equation}
\log P(\textrm{data}\mid \{\theta_i\}) = -\frac 1 2 \sum_i  \frac{\left[O_i(\mathrm{data})-O_i(\{\theta_i\})\right]^2}{\Delta O_i^2}  \, ,
\label{eq:likelihood}
\end{equation}
where the sum goes over the number of data indexed by $i$, where $O_i(\mathrm{data})$ represents a measured data and $O_i(\{\theta_i\})$ its prediction from the model. $\Delta O_i^2$ is related to the uncertainty in the data originating from the measurement, but it can also incorporate the uncertainty of the model~\cite{Dobaczewski:2014}. Finally, the probability $P(\{\theta_i\})$ in Eq.~\eqref{eq:bayes} is the prior probability, which represents the \textsl{a-priori} knowledge on the model parameters. 

The Bayesian framework allows us to investigate the propagation of experimental uncertainties on i) our model's parameters as well as on ii) our predictions. By marginalizing over all other parameters, one could generate a Probability Distribution Function (PDF) associated to the model parameters and to our predictions.

The fixed parameters of this model are given in Table~\ref{tab:fixed_parameters}. They are consistent with the previous applications of this model~\cite{Massot2008,Rahul2022,Chamseddine,Chamseddine_NJL}. At the Hartree level, the contributions are governed by the ratio $(g_i/m_i)$ for the quadratic interactions (as is the case for $i=\omega, \rho$ interactions), so we simply fix the masses to the bare ones and keep only the coupling constants as variables.
The code used for finite nuclei computations is from \cite{NIKSIC_code}, adapted to our current model.

\begin{table*}[t]
\tabcolsep=0.5cm
\def\arraystretch{1.5}
\caption{\label{tab:fixed_parameters}%
Model parameters (masses and coupling constants) fixed in the present analysis.}
\begin{tabular}{ccccc}
\hline \hline
$M_N$ & $m_{\omega}$ &  $m_{\rho}$ & $m_{\pi}$ & $f_{\pi}$ \\
MeV & MeV &  MeV & MeV & MeV \\
\hline
938.9 & 783.0 &  779.0 & 139.6 & 94.0\\
\hline \hline
\end{tabular}
\end{table*}

The variable model parameters are $m_s$, $g_s$, $C_\mathrm{NS}$, $g_\omega$ and $g_\rho$. They are fitted to reproduce the NEPs given in Table~\ref{tab:NEP_fit}, as well as the binding energies and the charge radii measured for the doubly magic nuclei given in Table~\ref{tab:FN_fit}.

\begin{table}[t]
\tabcolsep=0.33cm
\def\arraystretch{1.5}
\caption{
The NEPs (centroids and standard deviations), taken from \cite{Margueron2018}, which are imposed by adjusting the variable model parameters $m_s$, $g_s$, $C_\mathrm{NS}$, $g_\omega$ and $g_\rho$:
}
\begin{tabular}{ccc}
\hline\hline
NEP &  centroid & std. dev.\\
\hline
$E_{\sat}$ (MeV) & -15.8 & 0.3\\
$n_{\sat}$ (fm$^{-3}$) &  0.155 & 0.005\\ 
$E_\sym$ (MeV) &  32 & 2 \\
$K_\sat$ (MeV) & 230 & 20 \\
\hline\hline
\label{tab:NEP_fit}
\end{tabular}
\end{table}

\begin{table}[t]
\tabcolsep=0.33cm
\centering
\caption{Binding energies and available charge radii measurements of the seven doubly magic nuclei considered for the fit. The experimental error bars are given in parenthesis, and (-) stands
for experimental error-bars smaller than the number of digits shown in the table.}
\begin{tabular}{ccccc}
\hline\hline
$Z$ & $N$ & nucleus & $B$ (MeV) & $R_c$ (fm) \\
    &     &         & Ref.~\cite{binding_energies} & Ref.~\cite{charge_radii} \\
\hline
8  & 8   & $^{16}$O   & $-127.6193(-)$  & $2.6991(52)$ \\
20 & 20  & $^{40}$Ca  & $-342.0521(-)$  & $3.4776(19)$ \\
20 & 28  & $^{48}$Ca  & $-416.0009(1)$  & $3.4771(20)$ \\
28 & 28  & $^{56}$Ni  & $-483.9956(4)$  & - \\
50 & 50  & $^{100}$Sn & $-825.2944(3000) $  & -  \\
50 & 82  & $^{132}$Sn & $-1102.8430(20) $ & $4.7093(76)$ \\
82 & 126 & $^{208}$Pb & $-1636.4301(11)$ & $5.5012(13)$ \\
\hline\hline
\end{tabular}
\label{tab:FN_fit}
\end{table}

The uncertainties considered in the likelihood probability~\eqref{eq:likelihood} are now detailed. 
We consider the following uncertainty for the binding energy~\cite{Odilon2023}
\[
\Delta_B = 2.0 ~\text{MeV}.
\]
The charge radius is defined via the known empirical formula~\cite{Bender:2003}
\begin{equation}
R_c^2 =  R_p^2  + 0.64~\text{fm$^2$}~,
\end{equation}
where $R_p$ denotes the root mean square (rms) radius of the proton density distribution and the term 0.64 fm$^2$ approximately accounts for the finite size of the proton and the contribution of the spin-orbit and the neutron to the charge radius~\cite{NIKSIC_code}. We consider the following uncertainty for the charge radius~\cite{Odilon2023}
\[
\Delta_{R_c} = 0.1A^{-1/3}~\text{fm}~.
\]
No priors are imposed on the parameters, except that they are positive (to reduce the parameter space to be explored to only physical values).

Finally, we consider the best model corresponding to the parameter set for which the posterior probability is maximal, and we refer to it as RMF-CC1 from now on.

\section{Results}
\label{sec:results}

In the following, we present and discuss our results: we start with doubly magic nuclei, and then we consider open-shell nuclei.

\subsection{Doubly magic nuclei}
\label{sec:dmn}

\begin{figure}[t]
\centering
\includegraphics[width=0.98\textwidth]{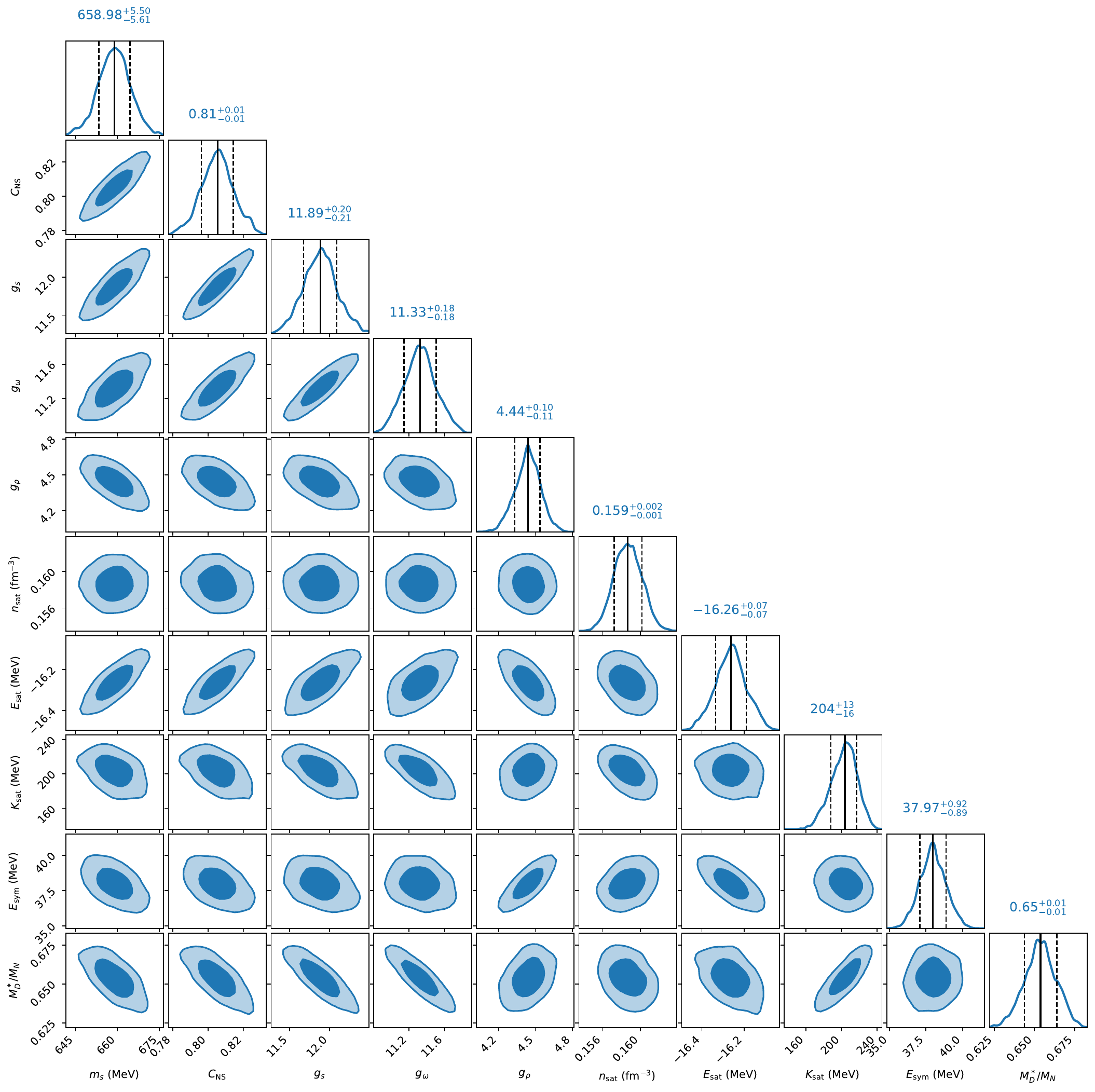}
\caption{Corner plot showing the joint (off-diagonal panels) and marginalized (diagonal panels) posterior distributions for the parameters ($m_s$ in MeV, $C_\mathrm{NS}, g_s, g_\omega, g_\rho$) and the NEPs: $n_\sat$ (fm$^{-3}$), $E_\sat, K_\sat, E_\sym$ in MeV, and the Dirac mass $M^*_D/M_N$ of the RMF-CC1 model fitted to the data of Tables~\ref{tab:NEP_fit} and \ref{tab:FN_fit}.}
\label{fig:corner_plot}
\end{figure}

After performing the fitting procedure discussed in the previous section, we are now ready to analyze the obtained results. Figure~\ref{fig:corner_plot} displays the PDFs for the model parameters and the NEPs, which allows us to extract the model uncertainty, defined as the $68\%$ credible interval (CI) of the distributions. We thus obtain the lower and upper bounds $\sigma_{\text{model}}^{\pm}$. The resulting optimal parameters (maximizing the posterior) obtained from the fitting, referred to as RMF-CC1, are shown in Table~\ref{tab:params_nopairing}, where we have also shown the parametrization from the well-known non-linear scalar interaction model NL3~\cite{Lalazissis1996} and the density-dependent DD-ME2~\cite{Lalazissis2005}. We additionally consider the Parity Doublet Model (PDM), applied to finite nuclei in Ref.~\cite{PDM_Kim} for 2 reasons: it also employs a chiral potential similar to our model, however they go to higher orders in the field (up to 6th order) with extra free parameters. The second reason is that the nucleon mass also depends on the scalar field, but the origin of this dependence is from the chiral partner of the nucleon, $N^*(1535)$, and not from confinement effects as in RMF-CC. In the following we consider the parameter set 3 of their paper, named PDM-3. This model turns out to be an important once we discuss pairing in Section~\ref{sec:pairing}.

The NL3 model shares several similarities with RMF-CC one, since both contain a non-linear scalar interaction. However, while the cubic and quartic coefficients are fixed by chiral symmetry in RMF-CC, see Sec.~\ref{sec:framework}, these coefficients are treated as free parameters in NL3. Additionally, the nucleon polarization in both NL3 and DD-ME2 is not considered, which can be reproduced in the RMF-CC approach by setting $C_\mathrm{NS}=0$. For a detailed comparison between RMF-CC and NL3, see Ref.~\cite{Rahul2022}.

\begin{table}[t]
\centering
\caption{Model parameters maximizing the posterior probability for the FN data and the NEPs (RMF-CC1). The quoted uncertainties correspond to 68\% credible intervals from the Bayesian posterior distribution. We additionally show the NL3~\cite{Lalazissis1996}, DD-ME2~\cite{Lalazissis2005} and PDM-3~\cite{PDM_Kim} models for comparison. The (-) indicates parameters undefined for the specific model, and the details of the parameterization can be found in~\cite{PDM_Kim}.}
\renewcommand{\arraystretch}{1.35}
\setlength{\tabcolsep}{6pt}
\begin{tabular}{lcccc}
\hline\hline
Parameter & RMF-CC1 & NL3 & DD-ME2 & PDM-3  \\
\hline
$m_s$ (MeV)             & 659.474$^{+5.496}_{-5.605}$ & 508.194 & 550.124 & 382.140 \\
$g_s$                   & 11.927$^{+0.204}_{-0.210}$ & 10.217 & 10.5396 & - \\
$g_\omega$              & 11.373$^{+0.182}_{-0.182}$ & 12.868 & 13.0189 & 7.036 \\
$g_\rho$                & 4.430$^{+0.097}_{-0.107}$ & 4.474 & 3.6836 & 3.958 \\
$C_\mathrm{NS}$         & 0.8064$^{+0.0089}_{-0.0092}$ & 0 & 0 & - \\
\hline\hline
\end{tabular}
\label{tab:params_nopairing}
\end{table}

The first quantities of interest are structure properties, mainly the binding energies and the charge radii. 
Note that this is the first application of the RMF-CC approach to finite nuclei and we would like to check whether our parametrization was able to correctly predict these nuclear structure properties, all the while respecting NEP constraints. 
Table~\ref{tab:NEP_predicted} shows the reproduced NEPs obtained for different models. We first notice the symmetry energy, which is larger than the experimental value by about 20\%. The corresponding slope parameter is
$L_{\sym}=114\pm3$ MeV, close to the NL3 value of $118$ MeV and significantly larger than the DD-ME2 and PDM-3 values of $51$ and $82$ MeV, respectively. The posterior favors a rather stiff symmetry energy despite the nominal data constraint.
A large value of $L_{\rm sym}$ implies a larger pressure in neutron-rich matter around saturation density and, in neutron-star applications, generally tends to favor larger stellar radii and tidal deformabilities. As discussed in
Ref.~\cite{Pradhan2026}, astrophysical constraints tend to disfavor such a stiff isovector sector in RMF-CC, indicating a tension between the present finite-nucleus/NEP
calibration and astrophysical constraints.
 This indicates a residual tension in the isovector channel of the present density-independent RMF-CC1, similar to what is observed in stiff nonlinear RMF parameterizations such as NL3. This suggests that a density-dependent isovector coupling, or an additional isovector degree of freedom, may be required for a quantitatively accurate isovector sector, as is the case for DD-ME2 for example.

The incompressibility modulus also deserves some discussion. The optimal RMF-CC1 value, $K_{\sat}=199^{+13}_{-16}$ MeV, lies below the centroid $230\pm20$ MeV used in the likelihood in Table~\ref{tab:NEP_fit}. The NEPs, however, enter the calibration as Gaussian constraints rather than as hard bounds, so the optimal parametrization
results is a compromise between all nuclear-matter and finite-nucleus observables and is not required to remain within the standard deviation of each individual
constraint. In the present model, the posterior favors a somewhat lower value of $K_{\sat}$. \\
In RMF-CC, the incompressibility is strongly constrained by the scalar sector, whose functional form is governed by the chiral potential and the nucleon response. The same limited set of scalar parameters must simultaneously reproduce the
saturation properties and the FN observables, leaving less freedom to adjust the curvature of the equation of state around saturation independently. The posterior therefore favors a parametrization with a somewhat lower $K_{\sat}$ in the global compromise. As shown later in Sec.~\ref{sec:systematics}, the strong correlation of $K_{\sat}$ with other quantities governed by the scalar sector, such as the Dirac effective mass, further illustrates this restricted flexibility.

Table~\ref{tab:NEP_predicted} also displays two "effective masses", the Dirac mass, which appears in the Dirac equation~\eqref{eq:Dirac_eqn} as
\begin{equation}
M^*_D=M_N + \Sigma_S~,
\end{equation}
known to be related to the spin-orbit potential strength, and the non-relativistic (NR) mass, $M^*_{NR}$, known to influence the density of single-particle states. The latter can be extracted from the NR reduction of the Dirac equation into a Schrodinger-equivalent equation (see Ref.~\cite{Chamseddine}), and is defined for nuclear matter at the Hartree level as:
\begin{equation}
M^*_{NR} = M_N - \Sigma_0~.
\end{equation}
It should be noted that these 2 masses are governed by different mesons, where $M^*_D$ is controlled by the scalar field $s$ while $M^*_{NR}$ by the vector fields $\omega$ and $\rho$. For more details on the different conventions on effective masses, see Refs.~\cite{Jaminon1989,Ma2004,vanDalen2005}.
Note that the Dirac and $NR$ masses are larger for both RMF-CC1 and PDM-3 compared to NL3 and DD-ME2. \\
In the following, all our comparisons of the RMF-CC1 (and PDM-3) outputs, namely the effective masses, were being done relative to these standard RMF parameterizations (NL3 and DD-ME2), which are used here as reference models for nuclear matter and FN properties. However, these comparisons are not an absolute characterization of the effective-mass values, since different calibration protocols can favor different ranges. RMF-CC parametrizations constrained by astrophysical observables can favor larger Dirac effective masses~\cite{Pradhan2026}.

\begin{table}[t]
\tabcolsep=0.33cm
\def\arraystretch{1.5}
\caption{
The reproduced NEPs using the parametrization of Table~\ref{tab:params_nopairing}, with the uncertainties corresponding to the 68\% credible intervals from the Bayesian posterior distribution, compared to the NL3~\cite{Lalazissis1996}, DD-ME2~\cite{Lalazissis2005} and PDM-3~\cite{PDM_Kim} predictions.
}
\begin{tabular}{ccccc}
\hline\hline
Parameters &  RMF-CC1 & NL3 & DD-ME2 & PDM-3 \\
\hline
$E_{\sat}$ (MeV) & -16.26$^{+0.07}_{-0.07}$ & -16.30 & -16.14 & -16.22 \\
$n_{\sat}$ (fm$^{-3}$) &  0.159$^{+0.002}_{-0.001}$ & 0.1483 & 0.152 & 0.16 \\ 
$E_\sym$ (MeV) &  38$^{+1}_{-1}$ & 37.4 & 32.30 & 30.85 \\
$K_\sat$ (MeV) & 199$^{+13}_{-16}$ & 272 & 251 & 215 \\
$L_\sym$ (MeV) & 114$^{+3}_{-3}$ & 118 & 51 & 82\\
$M^*_D / M_N$    & 0.65$^{+0.01}_{-0.01}$ & 0.60 & 0.57 & 0.83 \\
$M^*_{NR} / M_N$    & 0.72$^{+0.01}_{-0.01}$ & 0.67 & 0.65 & 0.89 \\
\hline\hline
\label{tab:NEP_predicted}
\end{tabular}
\end{table}

\begin{figure}[t]
\centering
\includegraphics[width=0.7\textwidth]{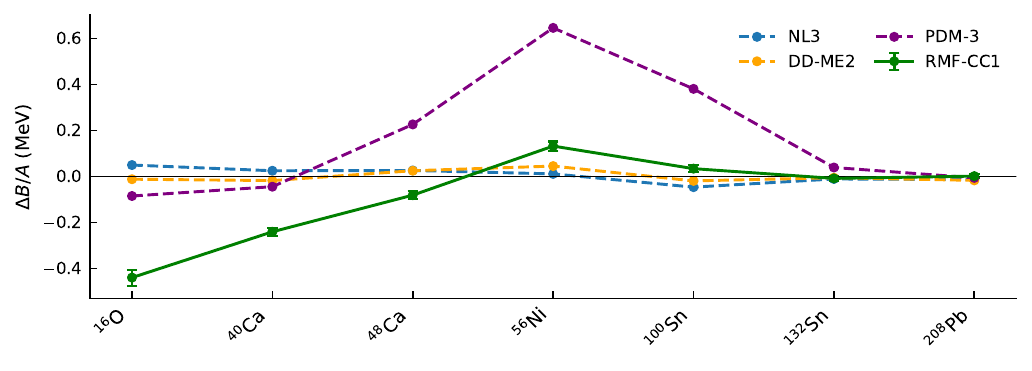}
\caption{Deviation of the theoretical binding energy per nucleon $\Delta B/A$~(MeV) from the experimental values of Table~\ref{tab:FN_fit}, using the parameterizations in Table~\ref{tab:params_nopairing}, for the RMF-CC1, NL3, DD-ME2 and PDM-3 models.}
\label{fig:B_magic}
\end{figure}

\begin{figure}[t]
\centering
\includegraphics[width=0.7\textwidth]{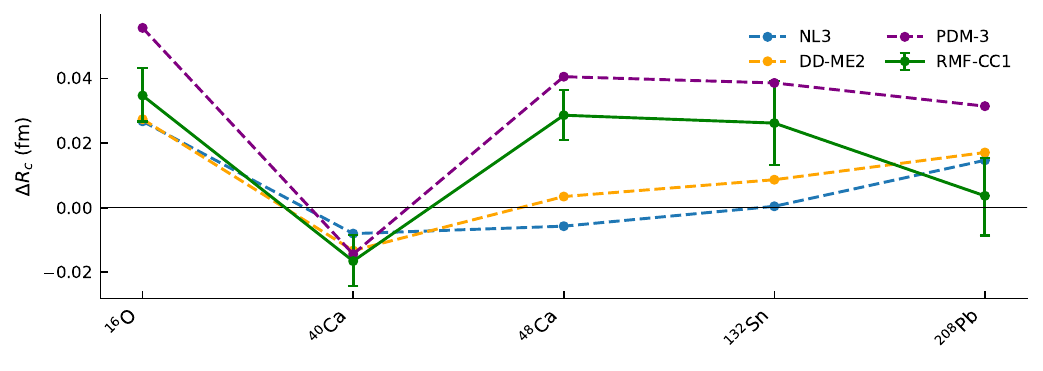}
\caption{Deviation of the theoretical charge radii $\Delta R_c$~(fm) from the experimental values of Table~\ref{tab:FN_fit}, using the parameterizations in Table~\ref{tab:params_nopairing}, for the RMF-CC1, NL3, DD-ME2 and PDM-3 models.}
\label{fig:rc_magic}
\end{figure}

\begin{figure}[t]
\centering
\includegraphics[width=0.7\textwidth]{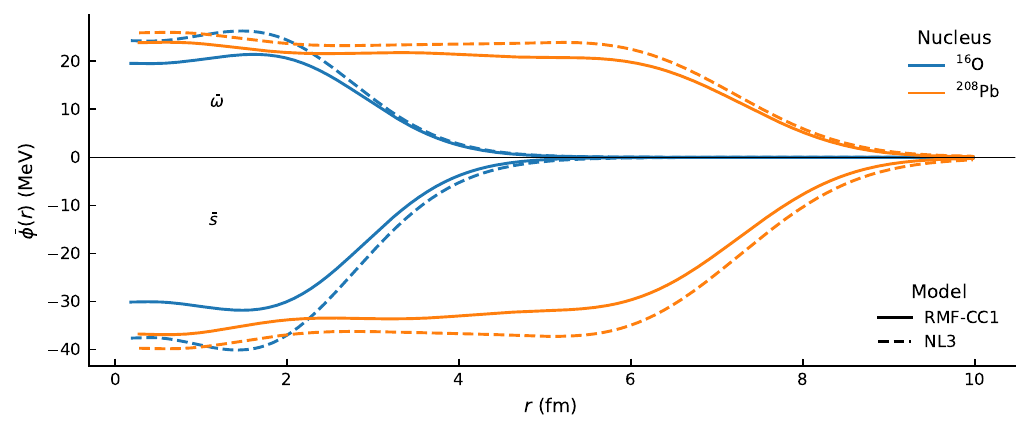}
\caption{Mean field profile of the scalar field $\bar s(r)$ and vector field $\bar \omega(r)$ as a function of the radius. We compare the profiles of a light nucleus $^{16}$O and a heavy one $^{208}$Pb for both RMF-CC1 and NL3 models. }
\label{fig:sigma_omega}
\end{figure}

The binding energies per nucleon and the charge radii of the nuclei employed in the fit are shown in Figs.~\ref{fig:B_magic} and \ref{fig:rc_magic}. The error bars are defined as $\sigma^{\pm} = \sqrt{\sigma_{\text{model}}^{\pm 2} + \sigma_{\text{exp}}^2} $. For the binding energy, we observe a bigger deviation from experiments for the light nuclei, especially $^{16}$O and $^{40}$Ca. In contrast, NL3 and DD-ME2 provide more flexibility through either free parameters in the non-linear scalar potential or density-dependent couplings, and are fitted to both light and heavy nuclei. In the RMF-CC model, the scalar potential is constrained by chiral symmetry, which is mainly calibrated through bulk properties. To see this, we plot the profile of $\bar s(r)$ and $\bar \omega(r)$ in Figure~\ref{fig:sigma_omega}, for both $^{16}$O and $^{208}$Pb, in RMF-CC1 and NL3, since they both exhibit non-linear scalar interactions. We observe no large qualitative differences in the field profiles, suggesting that the difference is not primarily due to the gradient terms associated with the meson fields themselves, but rather to the different functional dependence of the scalar contribution, in particular through the chiral potential and the nucleon polarization. The question we want to answer is why this difference plays a role for light nuclei and not heavy ones. For heavy nuclei, we have a large plateau for $\bar s,$ so the field remains nearly constant throughout the nucleus before quickly dropping toward zero at the surface. This means that the potential is mainly probed locally around this value $\bar s \approx -35$ MeV, and since both models are fitted to saturation properties, their potentials behave similarly locally. In contrast, light nuclei do not exhibit such a plateau, and therefore probe a broader region of the potential, and as we move away from saturation, the two potentials differ significantly (the cubic and quartic terms even carry opposite signs in RMF-CC1 and NL3, this is discussed in \cite{Rahul2022}). Since NL3 has free parameters in the scalar potential, it can more easily accommodate the properties of light nuclei than the RMF-CC1 potential, which remains strongly constrained by chiral symmetry.

Another possible source of this difference in density-dependence is the absence of explicit Fock terms.
Missing exchange contributions are effectively absorbed into the fitted Hartree couplings around the calibration domain. Away from saturation, and especially in light nuclei where surface and sub-saturation densities carry more weight, this absorption is no longer guaranteed to remain accurate, especially in RMF-CC, where the scalar potential is strongly constrained by chiral symmetry and therefore has less phenomenological flexibility.

For PDM-3, the unconstrained higher-order terms in the scalar-field potential may also provide additional flexibility in reproducing light nuclei through the fitting procedure, although a more detailed discussion is given in~\cite{PDM_Kim}.

\begin{figure}[t]
\centering
\includegraphics[width=0.7\textwidth]{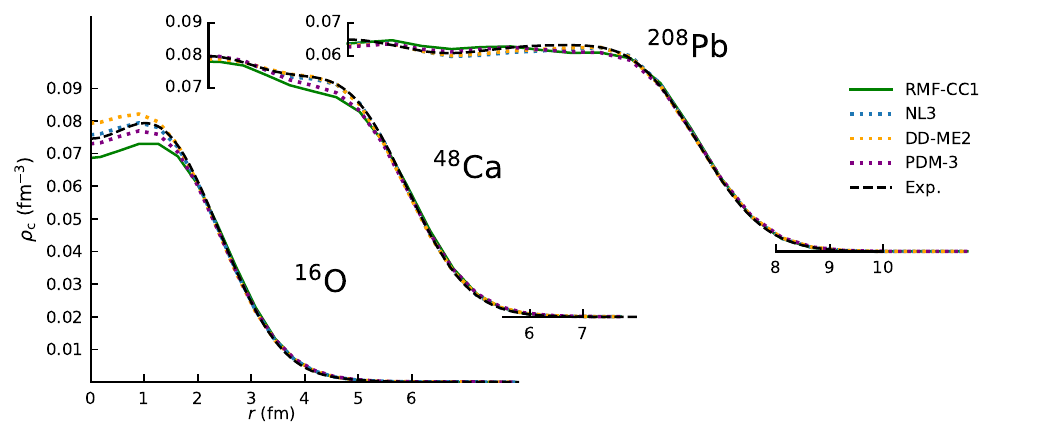}
\caption{Comparison of the charge density profile as a function of the radius, for the light $^{16}$O, medium $^{48}$Ca and the heavy $^{208}$Pb nuclei, for the three models RMF-CC1, NL3, DD-ME2 and PDM-3 and experimental data from~\cite{charge_density}.  }
\label{fig:charge_density}
\end{figure}

We can similarly look at the charge density profile in Fig.~\ref{fig:charge_density}, which is obtained after convoluting the proton density of Eq.~\eqref{eq:densities} with the proton form factor~\cite{Yao2012}. We can see it exhibits the same plateau behavior as we go from light to heavy nuclei, again reflecting the increasing importance of bulk properties in heavier nuclei. For $^{16}$O and $^{48}$Ca, the central density is reduced compared to other models, while the surface density is enhanced, shifting the charge outward and increasing the charge radius, as seen in Fig.~\ref{fig:rc_magic}. So RMF-CC1 exhibits a more diffuse charge density profile.

For the charge radii, the results between different models are quite similar, with less than $\sim 2\%$ deviation from experimental values, as seen in Figure~\ref{fig:rc_magic}.

\subsection{Open-shell nuclei}
\label{sec:pairing}

In this section, we consider the RMF-CC1 parameter set obtained from the fit to doubly magic nuclei and NEPs in Sec.~\ref{sec:dmn} and Table~\ref{tab:params_nopairing}, and we explore open-shell nuclei with a pairing interaction which is a Gogny-type separable interaction based on the Gogny force D1S~\cite{BERGER1991} rendering the computation easier~\cite{NIKSIC_code}. So far this pairing was turned off since we were considering closed shells only. 

\begin{figure}
\centering
\includegraphics[width=0.7\textwidth]{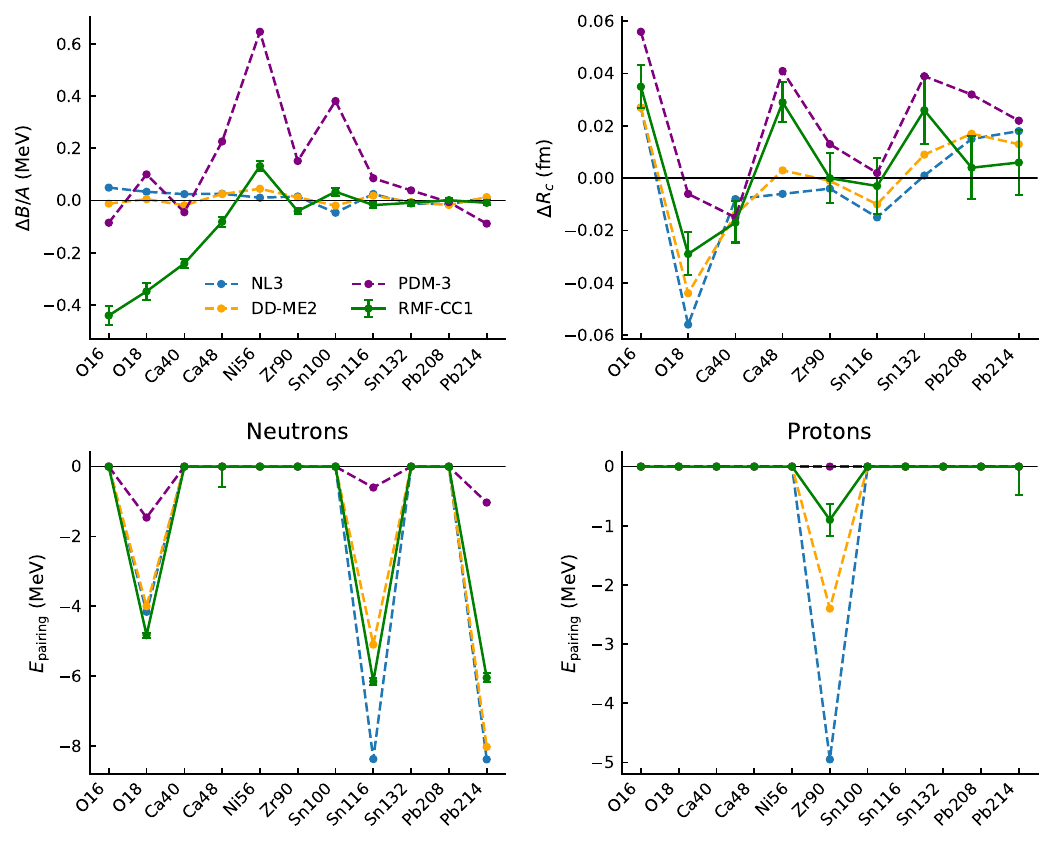}
\caption{Deviation of the theoretical binding energy per nucleon $\Delta B/A$~(MeV) and charge radii from the experimental values in \cite{binding_energies,charge_radii} (top panels), and comparison of pairing energies for neutrons/protons (bottom panels) using the parameterizations in Table~\ref{tab:params_nopairing}, for the RMF-CC1,NL3, DD-ME2 and PDM-3 models, with a separable Gogny force for the pairing interaction, after the reduction of the pairing strength.}
\label{fig:results_open}
\end{figure}

We observe a good description of heavy nuclei from RMF-CC1, while light and some intermediate mass nuclei are not well described, as previously observed.
The surprising part is the appearance of anomalous pairing in the case of the doubly magic nuclei previously studied for RMF-CC1. This is even more pronounced for the PDM-3. 
The pairing correlations from the separable Gogny force are not automatically zero for closed shells, in fact if the gap around the Fermi level is low enough, we could still have nonzero pairing. 
Additionally, to verify that this anomaly is not an artifact of the particular pairing interaction employed, we tested RMF-CC1 with three alternative pairing prescriptions: a zero-range delta interaction, a density-dependent delta interaction, and a full finite-range Gogny interaction. All three can also generate nonzero pairing correlations in  closed-shell nuclei, although the corresponding pairing energies depend on the interaction. This indicates that the anomalous pairing is not specific to the separable Gogny force and instead points to the underlying single-particle spectrum, in particular to small shell gaps around the Fermi energy.
The higher Dirac/NR effective mass in RMF-CC1 and PDM-3 compared to NL3 or DD-ME2, seen from Table~\ref{tab:NEP_predicted}, leads to a higher density of states around the Fermi level, which can explain why we might have this anomaly. Reducing the pairing strength for both protons and neutrons by $\sim 0.9$ is enough to remove this effect using the parametrization of Table~\ref{tab:params_nopairing} for RMF-CC1. However, PDM-3 remains largely anomalous and it had to be rescaled by $0.4$ in order to remove the anomalous pairing. The obtained results, after applying this reduced pairing strength, are shown in Figure~\ref{fig:results_open}.

We obtain a similar conclusion as before concerning the binding energies; the first couple of light nuclei have a stronger deviation from experiments for RMF-CC1, however, for intermediate mass to heavy nuclei, we have excellent agreement. The charge radii are still within $\sim 2\%$ deviation from the experimental values. We notice that even though the anomalous pairing is gone for the specific parameterization we considered from Table~\ref{tab:params_nopairing} after pairing strength reduction, posterior samples generated by the MCMC procedure still include parameterizations for which this anomaly persists, as can be seen from the error bars around the pairing energy for the closed neutron shell of $^{48}$Ca,~or the closed proton shell of $^{214}$Pb. In the PDM-3 case, with a NR effective mass of $M^*_{NR}=0.89M_N$, the anomaly is strongly suppressed by the strong reduction of the pairing strength, it is even zero for the open proton shell in $^{90}$Zr. Of course, this is a simplified pairing interaction, where the main focus is on structure properties like the binding energy and charge radius, where the effect of pairing energy is supposed to be negligible. The reduction of the pairing strength is more of a diagnostic treatment of pairing behavior, while a more systematic study is beyond the current scope of this work. For a spectroscopic study, a more careful treatment of pairing should be considered, for example, by refitting the pairing strength to some single-particle energy gaps. 

\begin{figure}[t]
\centering
\includegraphics[width=0.7\textwidth]{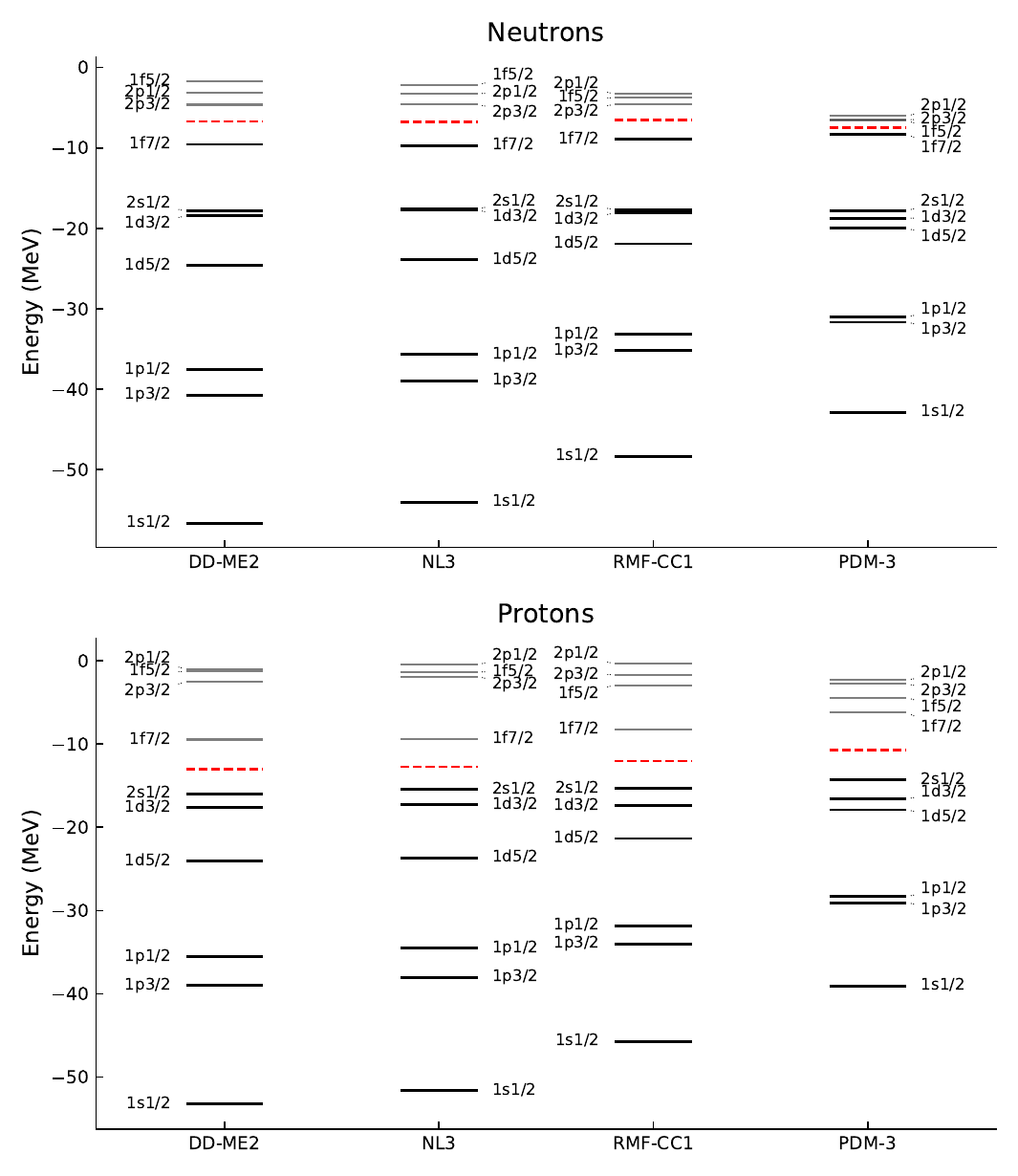}
\caption{Comparison of the single-particle energies of $^{48}$Ca between the various models, and for neutrons (top) and protons (bottom). The red line corresponds to the Fermi energy.}
\label{fig:spe}
\end{figure}

In Figure~\ref{fig:spe}, we show the single-particle energies (SPE) for the $^{48}$Ca nucleus, for both neutrons and protons. This nucleus was the first doubly magic nucleus to present an anomalous pairing, if not for the reduction of the pairing strength, and it was present in the neutron shell. The link to both Dirac masses and NR masses can now be clarified. This is directly visible in the $1p_{3/2}$–$1p_{1/2}$ and $1d_{5/2}$–$1d_{3/2}$ doublets, and is quantified in Table~\ref{tab:spe_neutrons} for the neutron SPE shell structure. For instance, the $p$-splitting decreases from about 3.2–3.3 MeV in DD-ME2/NL3 to 2.04 MeV in RMF-CC1 and 0.67 MeV in PDM-3, while the $d$-splitting is reduced from about 6.2 MeV to 3.82 MeV and 1.28 MeV, respectively. This reduced splitting is consistent with a weaker effective spin-orbit interaction and contributes to a compression of the shell structure around the Fermi energy. Consistently, the $N=28$ gap is also reduced, from about 5 MeV in DD-ME2/NL3 to 4.36 MeV in RMF-CC1 and as low as 1.75 MeV in PDM-3, indicating a much softer shell closure.

The empirical values reported in Table~\ref{tab:spe_neutrons} should, however, be interpreted with some caution. Single-particle energies (and thus spin-orbit splittings) are not directly observable quantities. They are inferred from experimental information, in particular from one-nucleon transfer reactions, where the strength associated with a given single-particle orbital can be distributed over several states (fragmentation). In Ref.~\cite{Ray1979}, for instance, empirical single-particle binding energies are obtained from spectroscopic-strength-weighted averages of the available transfer reaction data. The extracted values can therefore depend on the experimental information and analysis employed. This is illustrated by the significant spread among values reported in the literature; for example, the neutron $1d$ spin-orbit splitting in $^{48}$Ca obtained from the empirical energies of Ref.~\cite{Ray1979} is $8.92$ MeV, whereas Ref.~\cite{Long2007} presents an experimental value of $5.30$ MeV. The empirical values in Table~\ref{tab:spe_neutrons} therefore provide experimental guidance for the shell structure rather than precise quantitative benchmarks for the calculated single-particle spectrum.

A second observation is the modification of the level ordering above the Fermi energy. As shown in Table~\ref{tab:spe_neutrons}, RMF-CC1 and PDM-3 exhibit a negative value of $\Delta_{fp} = \epsilon(1f_{5/2}) - \epsilon(2p_{1/2})$, corresponding to an inversion of the $2p_{1/2}$ and $1f_{5/2}$ states compared to DD-ME2 and NL3. 
Since $1f5/2$ has a larger degeneracy, bringing it closer to the Fermi energy increases the effective level density in its vicinity, which enhances pairing correlations.

For $\Delta_{fp}$, the experimentally relevant information is primarily the relative ordering of the two particle levels, i.e. the sign of $\Delta_{fp}$, rather than its precise numerical value. Experimentally, these unoccupied neutron states of $^{48}$Ca can be accessed through neutron-addition reactions and appear as states of the neighboring $^{49}$Ca nucleus. Assuming that the relevant states in $^{49}$Ca correspond mainly to a neutron added to the $^{48}$Ca core, Ref.~\cite{MONTANARI2011} identifies the $p_{1/2}$ and $f_{5/2}$ single-particle neutron states at excitation energies of $\sim 2$ MeV and $\sim 4$ MeV, respectively. Associating these states with the $2p_{1/2}$ and $1f_{5/2}$ particle orbitals above the $N=28$ shell closure therefore indicates the  ordering as $2p_{1/2}<1f_{5/2}$, corresponding to $\Delta_{fp}>0$. RMF-CC1 and PDM-3 predict $\Delta_{fp}<0$ and hence an inversion of these two levels relative to the experimentally inferred ordering.

Taken together, the reduced spin-orbit splittings, the smaller $N=28$ gap, and the modified $2p$–$1f$ ordering indicate a compressed spectrum and an increased density of states around the Fermi energy in RMF-CC1 and PDM-3. These features correlate with the onset of anomalous pairing observed in the neutron shell of $^{48}$Ca.

These differences can be traced back to the larger effective masses obtained in RMF-CC1 and PDM-3, which are associated with their underlying Lagrangian structure (chiral potential + nucleon polarization). Larger values of the Dirac and NR effective masses lead to a denser single-particle spectrum and a reduced spin-orbit splitting, consistently implying stronger pairing.

\begin{figure}[t]
\centering
\includegraphics[width=0.7\textwidth]{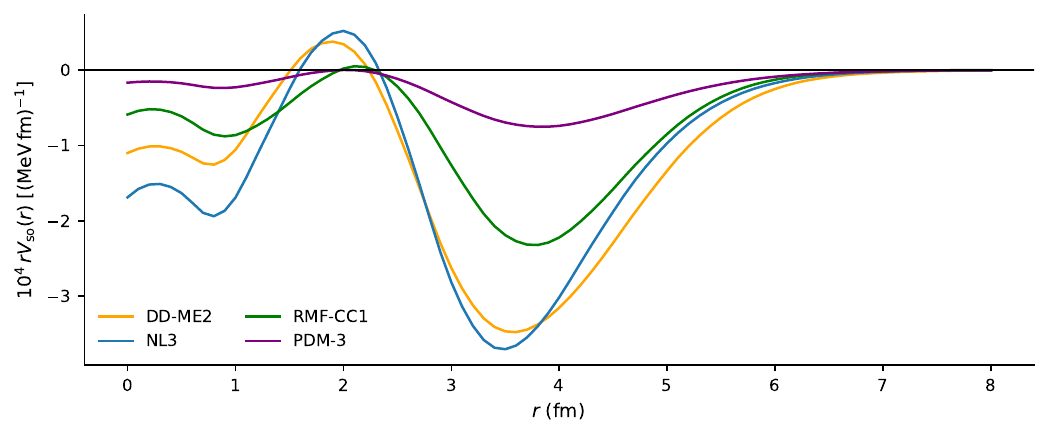}
\caption{Comparison of the radial dependence of the spin-orbit strength for neutrons in $^{48}$Ca for the DD-ME2, NL3, RMF-CC1 and PDM-3 models. The quantity $rV_{\mathrm{so}}(r)$ is shown to factor out the $1/r$ dependence in the spin-orbit potential $V_{\mathrm{so}}(r)$.}
\label{fig:spin-orbit}
\end{figure}

To finally conclude this discussion, we choose to represent the neutron's spin-orbit potential $V_{\mathrm{so}}(r)$ for $^{48}$Ca (we represent $rV_{\mathrm{so}}(r)$ to factor out the $1/r$ dependence in Eq.~\eqref{eq:spinorbit}). The formula is taken from~\cite{Ebrand-spinorbit} and given by

\begin{equation}
\label{eq:spinorbit}
V_{\mathrm{so}}(r) = \frac{1}{2r\,\mathcal{M}^2(r)} \frac{d}{dr}\bigl(\Sigma_0(r) - \Sigma_S(r)\bigr),
\end{equation}
with
\begin{equation}
\mathcal{M}(r) = M_N + \frac{1}{2}\bigl(\Sigma_S(r) - \Sigma_0(r)\bigr) = M^*_D(r) -\frac12\bigl(\Sigma_S(r) + \Sigma_0(r)\bigr)~.
\end{equation}
The plot is shown in Fig.~\ref{fig:spin-orbit}, for the 4 models, and it is fully consistent with our previous discussion. We notice that the spin-orbit strength is much weaker for RMF-CC1 and PDM-3, and the profile is much more diffuse, especially near the surface region at $r \sim 3.2$~fm, which is exactly what compresses the splitting of the spin-orbit doublets and leads to a higher density of states around the Fermi level.

A natural question raised by these results is why the RMF-CC1 model, which provides a satisfactory description of nuclear matter properties and bulk observables, exhibits a shell structure which is not supported by nuclear experimental data, in particular through a reduced spin–orbit splitting. This apparent tension reflects the fact that nuclear matter and finite nuclei probe different aspects of the underlying functional.

In nuclear matter, the constraints primarily fix the behavior of the model in a uniform medium, namely combinations of scalar and vector self-energies at a given density. In contrast, the spin–orbit interaction in finite nuclei is governed by the spatial structure of these fields, in particular their radial dependence in the surface region and the gradients of specific combinations of scalar and vector potentials. These quantities are not constrained by nuclear matter properties.

From this perspective, the reduced spin–orbit splittings, the compression of the single-particle spectrum, and the weakening of the N=28 shell gap observed in RMF-CC1 and PDM-3 can be interpreted as signatures of an effective spin–orbit potential that is too weak and/or too diffuse in the surface region, as we have seen in Fig.~\ref{fig:spin-orbit}. This naturally leads to a reduction of the gap around the Fermi energy, so a disappearance of magic numbers, and correlates with the appearance of anomalous pairing.

More generally, this behavior highlights a limitation of the RMF-CC framework: the strong constraints imposed by the chiral potential reduce the flexibility of the model to simultaneously reproduce bulk properties and detailed shell structure in finite nuclei. In contrast, phenomenological functionals such as NL3 or DD-ME2 can compensate this through free parameters, be it in the scalar potential (NL3) or in the density dependence of the couplings (DD-ME2).

\begin{table}[t]
\tabcolsep=0.25cm
\def\arraystretch{1.5}
\caption{
Neutron single-particle shell structure around the Fermi surface for $^{48}$Ca. 
The Fermi gap $\Delta_F$ indicates the energy gap between the first unoccupied and the last occupied levels. 
$\Delta_p = \epsilon(1p_{1/2})-\epsilon(1p_{3/2})$ and 
$\Delta_d = \epsilon(1d_{3/2})-\epsilon(1d_{5/2})$ denote spin-orbit splittings. 
$\Delta_{fp} = \epsilon(1f_{5/2})-\epsilon(2p_{1/2})$ indicates the relative ordering of the $1f_{5/2}$ and $2p_{1/2}$ levels (negative values correspond to inversion). 
Experimental values are extracted from Refs.~\cite{Ray1979,MONTANARI2011}. 
All energies are in MeV.
}
\begin{tabular}{cccccc}
\hline\hline
Quantity & DD-ME2 & NL3 & RMF-CC1 & PDM-3 & Exp. \\
\hline
$\Delta_F$ 
& $1f_{7/2}\!\to\!2p_{3/2}$ (4.93) 
& $1f_{7/2}\!\to\!2p_{3/2}$ (5.11) 
& $1f_{7/2}\!\to\!2p_{3/2}$ (4.36) 
& $1f_{7/2}\!\to\!1f_{5/2}$ (1.75)
& $1f_{7/2}\!\to\!2p_{3/2}$ (4.86)~\cite{Ray1979} \\

$\Delta_p$ 
& 3.19 & 3.30 & 2.04 & 0.67 & 4.30~\cite{Ray1979} \\

$\Delta_d$ 
& 6.18 & 6.16 & 3.82 & 1.28 & 8.92~\cite{Ray1979} \\

$\Delta_{fp}$ 
& +1.35 & +1.06 & $-0.47$ & $-0.55$ & +1.97~\cite{MONTANARI2011} \\
\hline\hline
\label{tab:spe_neutrons}
\end{tabular}
\end{table}

\section{Spin-orbit systematics within the RMF-CC1 posterior}
\label{sec:systematics}

In this section, we aim to answer the following question:
does the present RMF-CC framework allow one to simultaneously reproduce nuclear matter properties and a realistic spin-orbit strength in finite nuclei, or does it exhibit a structural tension imposed by the constrained form of the model?

More precisely, we want to determine whether the observed weak spin-orbit sector is due to a poor choice of parameters, or whether it reflects a limitation of the RMF-CC predictions, in which the same constrained degrees of freedom control both bulk properties and the spin-orbit behavior.

To answer this, we need to analyze the posterior distribution rather than individual parameterizations, and test whether there exist regions of the posterior that reconcile both nuclear matter constraints and spin-orbit properties.

We first identify a set of spin–orbit diagnostics and examine their correlations with the NEPs. Several quantities were computed: the $p$- and $d$-doublet splittings, denoted by $\Delta_p$ and $\Delta_d$, the gap around the Fermi energy $\Delta_F$, the maximum of the spin-orbit potential~\eqref{eq:spinorbit}, represented by the quantity $\max |rV_{\rm so}|$, and the integral measure $\int r^2 |V_{\rm so}|\,dr$. Since all these diagnostics display the same qualitative trends, we restrict the discussion to two representative quantities:
\begin{itemize}
\item the maximum spin-orbit strength $\max |rV_{\rm so}|$,
\item the $p$-doublet splitting $\Delta_p=\epsilon(1p_{1/2})-\epsilon(1p_{3/2})$.
\end{itemize}

\begin{figure}[t]
\centering
\includegraphics[width=\linewidth]{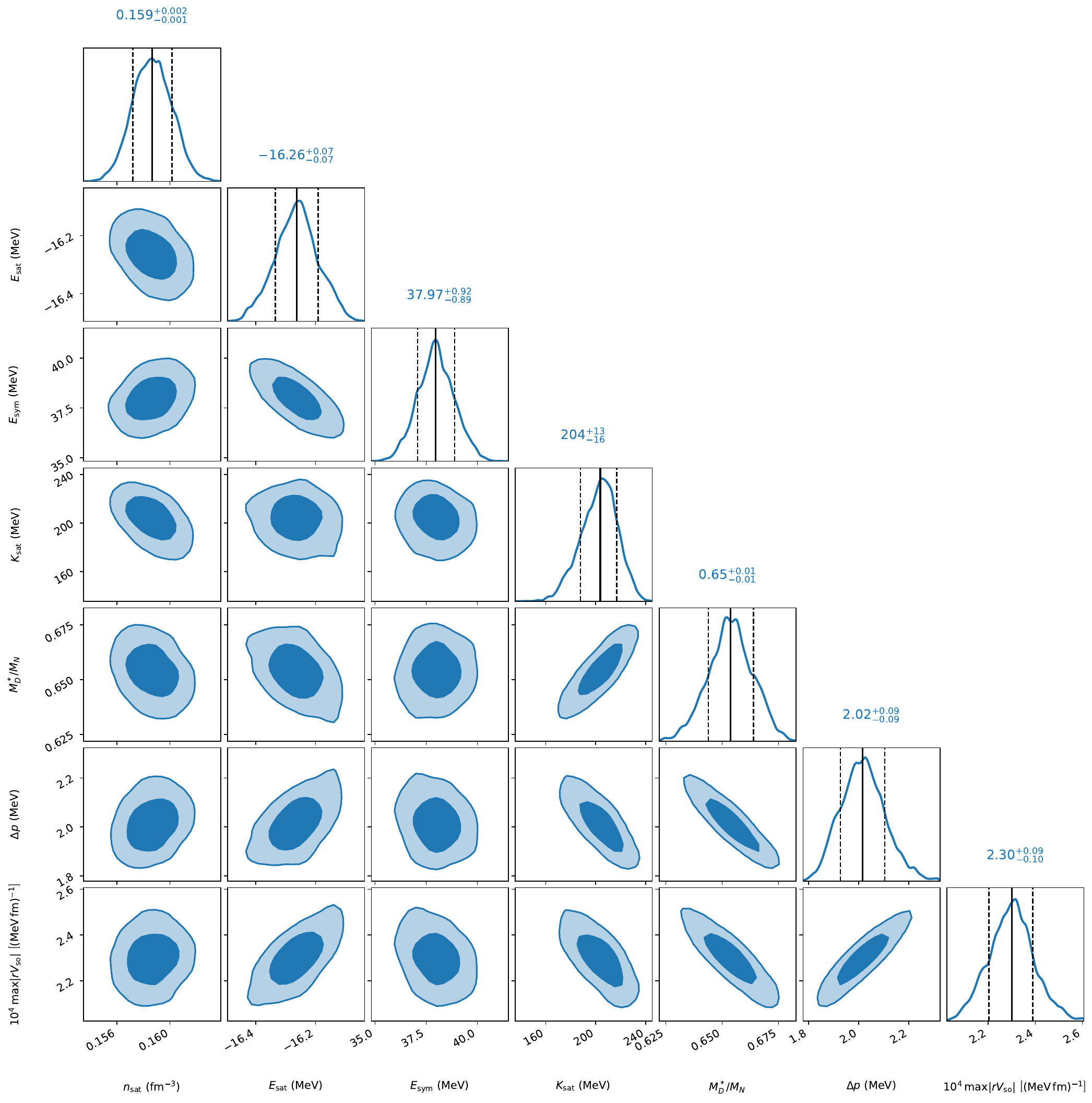}
\caption{Corner plot showing the joint and marginalized posterior distributions for the nuclear matter properties $n_{\sat}$, $E_{\sat}$, $E_{\sym}$, $K_{\sat}$, the Dirac mass $M_D^*/M_N$, and the spin-orbit diagnostics $\Delta_p$ and $\max |rV_{\rm so}|$. The posterior is obtained from the RMF-CC1 model fitted to the data of Tables~\ref{tab:NEP_fit} and~\ref{tab:FN_fit}.}
\label{fig:corner_NEP+so}
\end{figure}

The global correlations are shown in Fig.~\ref{fig:corner_NEP+so}, where the nuclear matter properties $n_{\sat}$, $E_{\sat}$, $E_{\sym}$, $K_{\sat}$, and the Dirac mass $M_D^*$ are plotted together with these two spin-orbit diagnostics.

A clear anti-correlation is observed between the Dirac mass and the spin-orbit diagnostics. However, this mass is also strongly anti-correlated with $K_{\sat}$; thus, lower values of $K_{\sat}$ are associated with larger $\max |rV_{\rm so}|$ and larger $\Delta_p$. \\
In the present RMF-CC framework, the spin-orbit potential, Dirac mass and the incompressibility $K_{\sat}$ are all controlled by the scalar channel, so it is expected that they are correlated. 

\begin{figure}[t]
\centering
\includegraphics[width=\linewidth]{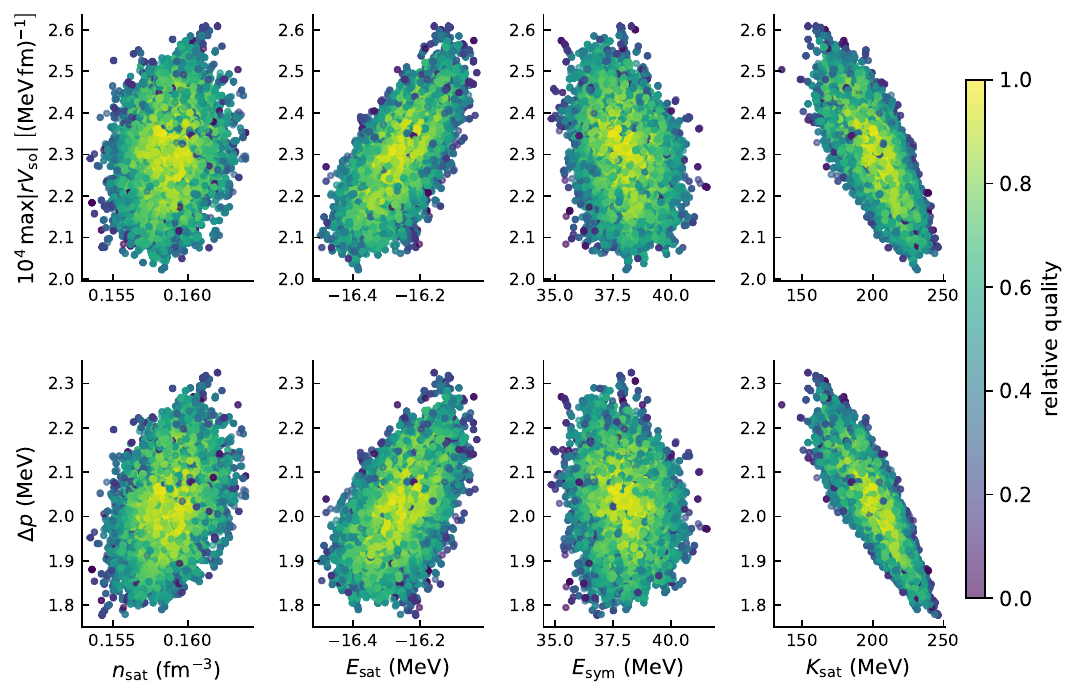}
\caption{Correlations between nuclear matter properties and the spin-orbit observables $\max |rV_{\rm so}|$ and $\Delta_p$. Each point corresponds to a posterior sample and is colored according to its relative posterior quality.}
\label{fig:so_vs_NEP}
\end{figure}

Now to explore whether there exists some good choice of parametrization to remedy the tension between NEPs and the spin-orbit sector, we plot in Fig.~\ref{fig:so_vs_NEP} the spin-orbit diagnostics against the nuclear matter properties, with each point colored according to its relative posterior quality,
\begin{equation}
q = \frac{\ell_p - \ell_p^{\min}}{\ell_p^{\max} - \ell_p^{\min}},
\end{equation}
where $\ell_p$ denotes the log-posterior value associated with a given parametrization, so that $q=1$ corresponds to the best posterior points.

The best posterior points remain in the bulk of the distribution, rather than moving toward the largest spin-orbit values. This indicates that increasing the spin-orbit strength within the present RMF-CC1 model requires moving away from the parameter region favored by the posterior.

\subsection{Achievable spin-orbit strength within the posterior}

To quantify the previous statement, we restrict the posterior sample to:
\begin{itemize}
\item the top 10\% of points in posterior quality,
\item an incompressibility window $K_{\sat}\in[210,250]$ MeV.
\end{itemize}
Within this subset, we extract for each spin-orbit diagnostic both the maximum value and the 95th percentile. The latter provides a more robust estimate of what is achievable without relying on extreme outliers. The results are compared to DD-ME2 and NL3 in Table~\ref{tab:so_conditional}.

\begin{table}[t]
\centering
\begin{tabular}{lcccc}
\hline \hline
Diagnostic & RMF-CC1 (max) & RMF-CC1 (95\%) & DD-ME2 & NL3 \\
\hline
$10^{4}\max |rV_{\rm so}|$ $(\mathrm{MeV\,fm})^{-1}$ 
& $2.31$ & $2.29$ & $3.47$ & $3.70$ \\
$\Delta p$ (MeV) 
& $2.01$ & $1.99$ & $3.19$ & $3.30$ \\
\hline \hline
\end{tabular}
\caption{Maximum and 95th percentile of the spin-orbit observables within the top 10\% posterior and the window $K_{\sat}\in[210,250]$ MeV, compared to DD-ME2 and NL3.}
\label{tab:so_conditional}
\end{table}

Two points are important. First, the maximum and the 95th percentile are very close, showing that there is no hidden posterior tail in which the spin-orbit strength is significantly higher. Second, even the largest values remain about $30$--$40\%$ below DD-ME2 and NL3. This favors the interpretation of a structural limitation of the present RMF-CC1 predictions, rather than a simple poor choice of parameter set.

\subsection{Origin of the weak spin-orbit}

To better understand the mechanism behind the reduced spin-orbit strength observed with RMF-CC1, and to determine whether it only originates from the increased effective mass or from surface effects, we decompose the spin-orbit potential in Eq.~\eqref{eq:spinorbit} as
\begin{equation}
rV_{\rm so}(r)=A(r)B(r),
\end{equation}
with
\begin{equation}
A(r)=\frac{d}{dr}(\Sigma_0-\Sigma_S), \qquad
B(r)=\frac{1}{2\mathcal{M}^2(r)}~.
\end{equation}
Thus, $A(r)$ measures the surface derivative of $\Sigma_0-\Sigma_S$, while $B(r)$ contains the effective-mass denominator.

We evaluate $A$ and $B$ at the radius $r^*$ where $|rV_{\rm so}|$ is maximal in the surface region. For RMF-CC1, three representative points are selected from the full posterior distribution, corresponding to the 10th, 50th, and 90th percentiles of $K_{\sat}$. The same decomposition is performed for DD-ME2 and NL3 using the identical definition of $r^*$. The results are shown in Table~\ref{tab:decomposition}.

\begin{table}[t]
\centering
\begin{tabular}{lccccc}
\hline \hline
Model & $K_{\sat}$ (MeV) & $r^*$ (fm) & $|A(r^*)|$ (MeV fm$^{-1}$) & $B(r^*)$ (MeV$^{-2}$) & $10^{4}\max |rV_{\rm so}|$ $(\mathrm{MeV\,fm})^{-1}$ \\
\hline
RMF-CC1 (low $K_{\sat}$)  & 183 & 3.8 & $2.74\times 10^2$ & $8.70\times 10^{-7}$ & $2.39$ \\
RMF-CC1 (mid $K_{\sat}$)  & 204 & 3.8 & $2.61\times 10^2$ & $8.53\times 10^{-7}$ & $2.23$ \\
RMF-CC1 (high $K_{\sat}$) & 220 & 3.8 & $2.58\times 10^2$ & $8.46\times 10^{-7}$ & $2.18$ \\
DD-ME2                      & 251 & 3.6 & $3.06\times 10^2$ & $1.14\times 10^{-6}$ & $3.47$ \\
NL3                         & 272 & 3.5 & $3.36\times 10^2$ & $1.10\times 10^{-6}$ & $3.70$ \\
\hline \hline
\end{tabular}
\caption{Decomposition of the spin-orbit potential into the surface factor $A(r)=d(\Sigma_0-\Sigma_S)/dr$ and the effective-mass factor $B(r)=1/(2\mathcal{M}^2(r))$, evaluated at the radius $r^*$ where $|rV_{\rm so}|$ is maximal.}
\label{tab:decomposition}
\end{table}

As discussed before, decreasing $K_{\sat}$ within the RMF-CC approach increases the spin-orbit strength. This variation is mainly driven by the surface term $|A(r^*)|$; as we move from the high to low $K_{\sat}$ value, $|A(r^*)|$ increases by about $6\%$, whereas $B(r^*)$ increases by about $3\%$. By looking at the variation of $\max |rV_{\rm so}|$, $|A(r^*)|$ makes up around $\sim 70\%$ of its variation. Thus, the variation of the spin-orbit strength is mostly governed by the surface derivative of $V-S$.

The comparison with DD-ME2 and NL3 clarifies the origin of the reduced spin-orbit strength. RMF-CC1 exhibits a smaller surface derivative than both DD-ME2 and NL3, and its effective mass factor $B(r^*)$ is also lower. As a result, the suppression of the spin-orbit strength cannot be attributed to a single mechanism. We saw that within the RMF-CC1 posterior, the  variation is mainly driven by the surface term $A(r^*)$. However, when compared to standard covariant EDFs, both contributions play a role. Relative to DD-ME2, the difference is indeed dominated by the effective mass factor, whereas for NL3 the surface and effective mass contributions are of similar importance. Overall, the reduced spin-orbit strength reflects the combined effect of a weaker surface derivative and a smaller effective mass factor.

Overall, these results suggest that the constrained scalar sector of the present RMF-CC model ties together the bulk curvature around saturation, the Dirac mass, and the surface derivative of $\Sigma_0-\Sigma_S$ more strongly than in the typical phenomenological covariant EDFs. This strong constraint, and the inability to reproduce the spin–orbit strength typical of standard covariant EDFs within an acceptable $K_{\sat}$ window points towards a missing surface/spin–orbit degree of freedom rather than to a mere readjustment of the nuclear-matter parameters. 

A natural way to restore this flexibility, which was considered in Ref.~\cite{Mercier-spinorbit}, would be to introduce an additional degree of freedom that is weakly constrained in uniform matter but active in finite nuclei, such as the $\omega$ tensor coupling, which is zero in infinite matter but directly modifies the spin-orbit in finite nuclei via the gradients of the $\omega$ field. Or, more microscopically, we can consider contributions beyond the Hartree level. In particular, the $\rho$ tensor channel can significantly enhance the spin–orbit strength, as discussed in Ref.\cite{Chanfray_rho}, and also reduce the Dirac and NR effective masses as shown in Ref.~\cite{Chamseddine}.

\section{Departing from the Linear Sigma Model}
\label{sec:NJL}

In this last section, we discuss the role of the scalar potential in our results.
So far, we considered the L$\sigma$M for our chiral potential~\eqref{eq:LsM}. This potential's cubic and quartic terms are fixed by chiral symmetry as mentioned before, however we can allow for a controlled departure from this fixed value to see its effect on the NEPs and finite nuclei properties. Phenomenologically, this can be motivated by models such as the NJL model, where one can generate a chiral potential based on a more accurate description of the low-energy realization of chiral symmetry breaking in the hadronic sector (see Refs.~\cite{Chanfray_Cchi,Chanfray-Schuck,Chamseddine_NJL}). As discussed in Ref.~\cite{Chanfray_Cchi,Chamseddine_NJL}, linearization of the NJL potential near saturation recovers the L$\sigma$M but modifies the cubic and quartic coefficients. The scalar potential can be expressed as
%given by
\begin{equation}
V_{\chi,\mathrm{NJL}}(s)\approx  \frac{1}{2}m^2_s {s}^2 +\frac{1}{2}\frac{m^2_s -m^2_\pi}{ f_\pi} \big(1-C_{\chi,\mathrm{NJL}}\big){s}^3 
+\frac{1}{8}\frac{m^2_s -m^2_\pi}{ f_\pi^2} \big(1-6~C_{\chi,\mathrm{NJL}} + D_{\chi,\mathrm{NJL}}\big) {s}^4
+...,
\label{eq:vchiNJL}
\end{equation}
where $C_{\chi,\mathrm{NJL}}$ and $D_{\chi,\mathrm{NJL}}$ are related to NJL loop integrals~\cite{Chanfray_Cchi,Chamseddine_NJL}. We see that for $C_{\chi,\mathrm{NJL}}=D_{\chi,\mathrm{NJL}}=0$ we simply recover the L$\sigma$M, while choosing $C_{\chi,\mathrm{NJL}}=1$ quenches the cubic term, as in the case of the QMC model~\cite{Guichon1988,Guichon2004}. These parameters have ranges $C_{\chi,\mathrm{NJL}} \in [0,1]$ and $D_{\chi,\mathrm{NJL}} \in [0,5]$. We will thus repeat our previous fitting procedure in Section~\ref{sec:parametrisation} while allowing the cubic and quartic coefficients to vary in their respective intervals by imposing it as a prior. Technically, these 2 parameters depend on the same underlying set of hadronic parameters, and so are not independent, but for the sake of this qualitative study, we will treat them as such. The new parametrization is given in Table~\ref{tab:params_NJL}.

\begin{table}[t]
\centering
\caption{Model parameters for RMF-CC maximising the posterior probability for the FN data and the NEPs, for the L$\sigma$M and NJL potentials. }
\begin{tabular}{lcc}
\hline\hline
Parameter & L$\sigma$M & NJL \\
\hline
$m_s$ (MeV)             & 659.474 & 564.225 \\
$g_s$                   & 11.9268 & 11.3178 \\
$g_\omega$              & 11.3726 & 12.9789 \\
$g_\rho$                & 4.4302 & 4.2592 \\
$C_\mathrm{NS}$         & 0.8064 & 0.2537 \\
$C_{\chi,\mathrm{NJL}}$         & 0 & 0.9588 \\
$D_{\chi,\mathrm{NJL}}$         & 0 & 3.4985 \\
\hline\hline
\end{tabular}
\label{tab:params_NJL}
\end{table}

We comment first on the values of $C_\mathrm{NS}$, $C_{\chi,\mathrm{NJL}}$ and $D_{\chi,\mathrm{NJL}}$ obtained with the NJL potential. The decrease of $C_\mathrm{NS}$ compared with the L$\sigma$M case is actually not surprising.
As discussed in Ref.~\cite{Chanfray_Cchi}, the nucleon response $C_\mathrm{NS}$ generates a repulsive three-nucleon contribution in the scalar sector, which plays an important role in the saturation mechanism.
In the L$\sigma$M, this repulsion has to compensate the attractive three-nucleon contribution associated with the cubic term of the chiral potential. With the NJL  potential, the parameter
$C_{\chi,\mathrm{NJL}}$ reduces this attractive contribution, making the scalar three-nucleon interaction more repulsive. A smaller value of $C_\mathrm{NS}$ is therefore required to obtain a similar three-nucleon contribution.

As shown in Ref.~\cite{Chanfray_Cchi}, the relevant quantity to compare between the two scalar potentials is therefore the effective combination
\begin{equation}
    \widetilde{C}_3 = \frac{M_N}{g_s F_\pi}C_\mathrm{NS} + \frac{C_{\chi}}{2},
\end{equation}
which governs the repulsive three-nucleon contribution.
Using the parameters of Table~\ref{tab:params_NJL}, we obtain $\widetilde{C}_3\simeq0.68$ for the L$\sigma$M potential and $\widetilde{C}_3\simeq0.70$ for the NJL potential. The two values are therefore very similar, despite the much smaller fitted value of $C_\mathrm{NS}$ in the NJL case.

According to the recent QCD/FCM-inspired studies of Refs.~\cite{Chanfray_CCM,chanfray2026}, $C_\mathrm{NS}$ is expected to lie approximately in the range $C_\mathrm{NS}\sim0.3$--$0.45$, while a value of order $C_\mathrm{NS}\sim0.5$ is expected in the MIT-bag realization of the QMC model. On the other hand, usual NJL parameter sets typically lead to $C_{\chi,\mathrm{NJL}}$ of order $0.5$~\cite{Chanfray_Cchi}.
The individual values of $C_\mathrm{NS}$ and $C_{\chi,\mathrm{NJL}}$ obtained in the present fit may therefore differ from these microscopic expectations, even though the effective combination $\widetilde{C}_3$ governing the scalar
three-nucleon contribution is very similar for the L$\sigma$M and NJL-inspired fits.

The value $C_{\chi,\mathrm{NJL}}\simeq0.96$ obtained here is close to the upper limit considered in our analysis, and it would be difficult to obtain with usual microscopic NJL parameter sets had we done a systematic NJL study. The purpose of the present analysis is, however, not to perform a consistent microscopic NJL calculation, but rather to use the NJL-inspired form of the potential to study how variations of the cubic and quartic terms of the
chiral potential affect FN properties. A consistent NJL study would require relating $C_{\chi,\mathrm{NJL}}$ and $D_{\chi,\mathrm{NJL}}$ to the underlying microscopic parameters, as done in Refs.~\cite{Chamseddine_NJL,Chanfray_Cchi}.

\begin{figure}[t]
\centering
\includegraphics[width=0.7\textwidth]{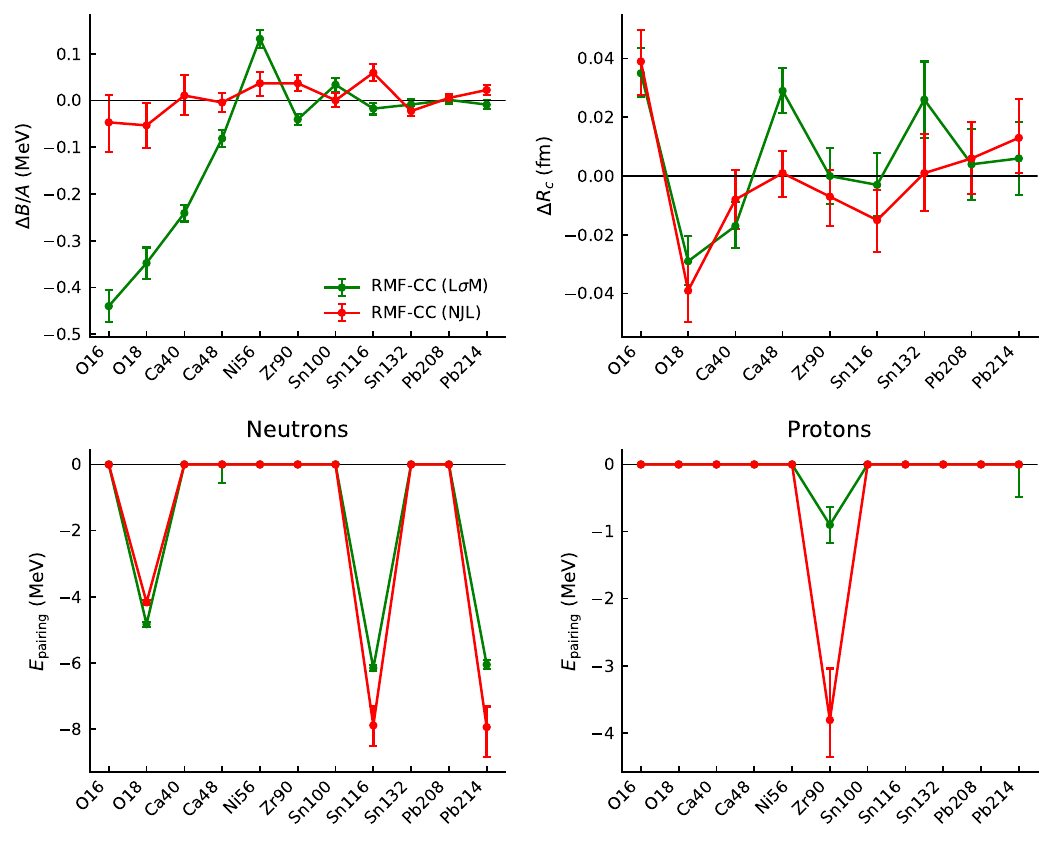}
\caption{Deviation of the theoretical binding energy per nucleon $\Delta B/A$~(MeV) and charge radii from the experimental values from Refs.\cite{binding_energies,charge_radii} (top panels), and pairing energies comparison for neutrons/protons (bottom panels) using the parameterizations in Table~\ref{tab:params_NJL}, for the RMF-CC with L$\sigma$M and NJL potentials, with a separable Gogny force for the pairing interaction, after the reduction of the pairing strength. }
\label{fig:results_open_NJL}
\end{figure}

We move on to present the NEP results in Table~\ref{tab:NEP_NJL}. We observe a slight lowering of the symmetry energy, while the incompressibility modulus is now within expected experimental window. However, the main point is the Dirac and NR effective masses, which are now lowered down to 0.58$M_N$ and 0.66$M_N$ respectively. This is expected to enhance the spin-orbit strength and open the shell gaps around the Fermi surface, and the previously observed pairing anomaly should vanish. Indeed, results in Figure~\ref{fig:results_open_NJL}, 
show an improvement of the predictions for the binding energy and charge radii for intermediate and low mass nuclei. In addition, the pairing anomaly also disappears.
This is detected by the fact that
all parameterizations within the 68\% CI for the NJL potential give zero pairing at closed shells, unlike the L$\sigma$M ones, where we still observed some small but non-zero pairing for $^{48}$Ca (neutrons) and $^{214}$Pb (protons). This confirms our previous justification about the importance of the potential for light nuclei: by relaxing the strongly constrained chiral potential, we are able to better reproduce the properties of light and intermediate mass nuclei. Of course, this relaxation comes with 2 new parameters to fit, similarly to NL3. However, these 2 parameters are constrained to remain inside a certain range allowed by the NJL model. This also confirms the role of the Dirac and NR effective masses in the 
shell structure.
This is, however, a qualitative result; it allows us to better 
understand the role of the chiral potential and the Dirac mass emerging from RMF-CC and PDM-3. 
The former seems to affect light nuclei, and the latter strongly affects the shell structure.

\begin{table}[t]
\tabcolsep=0.33cm
\def\arraystretch{1.5}
\caption{
The reproduced NEPs using the parametrization of Table~\ref{tab:params_NJL}. We show both the L$\sigma$M and the NJL predictions.
}
\begin{tabular}{ccc}
\hline\hline
Parameters &  L$\sigma$M & NJL \\
\hline
$E_{\sat}$ (MeV) & -16.26 & -16.10 \\
$n_{\sat}$ (fm$^{-3}$) &  0.159 & 0.152 \\ 
$E_\sym$ (MeV) &  38 & 36.98 \\
$K_\sat$ (MeV) & 199 & 234 \\
$M^*_D / M_N$    & 0.65 & 0.58 \\
$M^*_{NR} / M_N$    & 0.72 & 0.66 \\
\hline\hline
\label{tab:NEP_NJL}
\end{tabular}
\end{table}

\begin{figure}[t]
\centering
\includegraphics[width=0.7\textwidth]{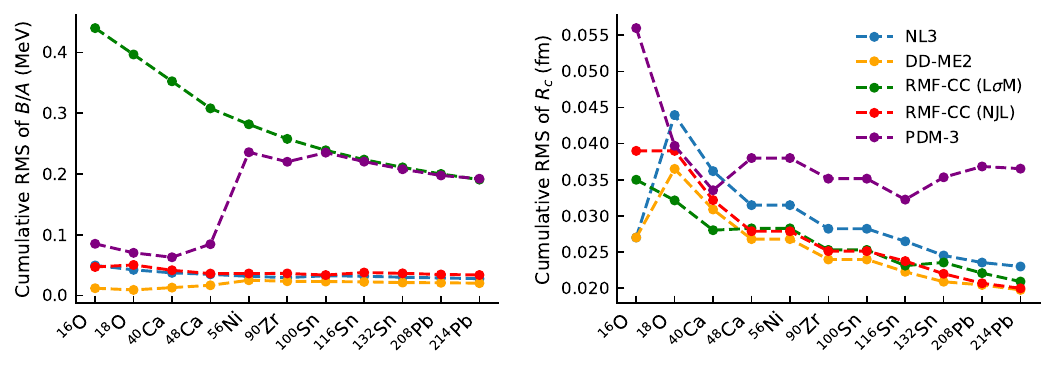}
\caption{The cumulative root mean square for the various models considered in this work for the binding energy per nucleon $B/A$ (MeV) and the charge radius $R_c$ (fm).}
\label{fig:rms}
\end{figure}

Finally, Figure~\ref{fig:rms} shows the cumulative root mean square (rms) for the various models for $B/A$ and $R_c$, which allows us to have a test on the predictions of these various models as we move from lighter to heavier nuclei. 
As discussed in Section~\ref{sec:results}, the RMF-CC model with the L$\sigma$M predictions improved when moving towards heavier nuclei. This trend can be seen as the rms gets smaller and smaller, reaching $\sim 0.2$~MeV. The PDM-3 model tends to the same final rms, however, the trend is that its predictions are degraded when we move to heavier nuclei. This rms is, however, one order of magnitude larger than the other models (NL3, DD-ME2), while the RMF-CC (NJL) predictions are on par with them, with an rms of $\sim 0.03 $ MeV.\\
In the case of charge radii, the RMF-CC model performs quite well with both versions of the potential, with a final rms of $\sim 0.02$ fm, an even better rms than that of the NL3 model.

\section{Conclusions}
\label{sec:conclusions}

In this work, we have considered an RMF framework supplemented with a chiral potential and a confinement-inspired nucleon response,, which was previously applied to nuclear matter~\cite{Massot2008,Rahul2022,Chamseddine,Chamseddine_NJL}. It has been extended to the description of finite spherical nuclei at the Hartree approximation for the first time.
The parameters of the model are adjusted to reproduce both NEPs at saturation and finite nuclei properties, such as binding energies and charge radii of doubly magic nuclei, and give the RMF-CC1 parametrization.
The resulting functional provides a consistent description of bulk nuclear properties while being a QCD-motivated covariant EDF: its scalar sector is constrained by chiral symmetry breaking through the identification of the scalar field with the chiral-invariant field associated with the radial fluctuation of the chiral condensate, while confinement-inspired nucleon response is encoded through a scalar polarizability introduced in Ref.~\cite{Chanfray2005} and associated with the self-consistent readjustment of the quark wave functions as originally proposed in Ref.~\cite{Guichon1988}.
The resulting parametrization provides a consistent description of bulk nuclear properties, while satisfying QCD-motivated constraints: the chiral potential is fixed and the identification of the scalar field with the chiral-invariant field associated with the radial fluctuation of the chiral condensate, and the inclusion of the nucleon response motivated by Ref.~\cite{Guichon1988}.

For doubly magic nuclei, we observed that the RMF-CC1  reproduces the binding energies and charge radii reasonably well within the considered uncertainties. The charge radii remain close to experiment, with less than $2\%$ deviation from experimental values, and comparable in quality to NL3 and DD-ME2, while the binding energies showed a more interesting trend: the largest deviations ($\sim 5\%$), occur for light nuclei, whereas the agreement improves significantly from medium to heavy nuclei. To understand the role of the scalar potential, we focused our attention on the RMF-CC1 and NL3 models, which share a non-linear potential, where the NL3 potential, however, is not constrained by chiral symmetry breaking. By looking at the scalar field profile inside the nucleus, we saw that the scalar field has a large plateau for heavier nuclei, so the system probes the scalar potential mostly around the saturation value, where both RMF-CC1 and NL3 potentials are fitted to behave similarly. However, for light nuclei, a broader interval of the potential is probed within the nuclear radius, making them more sensitive to the global shape of the scalar potential away from saturation. Since the RMF-CC potential is strongly constrained by chiral symmetry, it is less flexible than the phenomenological NL3 potential, whose nonlinear coefficients are adjusted directly to finite nucleus data. This naturally explains why NL3 performs better for light nuclei, whereas RMF-CC1 becomes more accurate for heavier systems, dominated by bulk properties. 

We then proceeded to study open-shell nuclei, equipped with a separable Gogny-type pairing interaction. We found that for the pairing strength considered for the force D1S~\cite{BERGER1991}, RMF-CC1 and PDM-3 exhibit an anomalous pairing in some closed-shell nuclei. This effect can be traced to the larger Dirac and NR effective masses predicted by these 2 classes of models, which reduce spin-orbit splittings and compress the single-particle spectrum around the Fermi surface and thus enhance pairing. A modest reduction of the pairing strength suppresses this anomaly at the diagnostic level for RMF-CC1, which had a Dirac (NR) effective mass of 0.65$M_N$ (0.72$M_N$), and gives a reasonable description of both closed and open-shell nuclei. However, it was not enough for the PDM-3 with a Dirac (NR) effective mass of 0.83$M_N$(0.89$M_N$), which required a drastic reduction in the pairing strength. We observe overall good agreement for medium and heavy nuclei, while again the RMF-CC1 has larger deviations compared to the NL3 and DD-ME2 models. To observe the effect of these effective masses, we looked at the SPE of the $^{48}$Ca, and observed both a reduction in the spin-orbit splitting, linked to the large Dirac mass compared to NL3 and DD-ME2,, which also changes the ordering of some energy levels above the Fermi level ($2p1/2$ and $1f5/2$) compared to NL3 and DD-ME2, and 
a reduction of the shell gaps around the Fermi level. 
Both of these effects can explain the observation of an enhanced and sometimes anomalous pairing. We also checked the spin-orbit potential and indeed confirmed that it is much weaker and more diffuse for RMF-CC1 and PDM-3, which have larger Dirac effective masses than NL3 and DD-ME2, again reinforcing the link with the weak spin-orbit splitting and the higher density of states around the Fermi level.

A systematic study was then done to answer two questions: do there exist within our RMF-CC1 phase space, parameterizations that are able to remedy the weak spin-orbit potential while respecting NEPs, and whether the major reduction of the spin-orbit potential is due to surface effects or merely the large effective mass. We managed to show that even the largest spin-orbit magnitudes reached are still 30-40\% lower than typical covariant EDFs such as DD-ME2 and NL3. Additionally, we traced back the origin of this deficit to an interplay between surface effects and the large effective mass exhibited by RMF-CC1.

Finally, we allowed for a small deviation away from the L$\sigma$M to explore the role played by the potential in the RMF-CC model. It supports that the global trend of the potential matters more for light nuclei, while the chiral potential performs very well for medium and heavy nuclei. It also confirms once again that the Dirac and NR effective masses are heavily responsible for the enhanced and thus anomalous pairing that is seen for some parameterizations, at least within the separable Gogny interaction considered here.

Overall, the present results show that RMF-CC constitutes a valid framework for the description of finite nuclei and nuclear matter, while incorporating QCD-motivated ingredients through the chiral structure of the scalar sector and a confinement-inspired nucleon response. However, there is a limitation appearing when considering a more systematic study for pairing and when trying to have a more global fitting procedure: the effective masses favored by the present finite-nucleus and NEP calibration are larger than in the NL3 and DD-ME2 reference models, contributing to a weak spin-orbit potential and an enhanced pairing interaction. The strongly constrained scalar sector of RMF-CC ties together the bulk curvature around saturation, the Dirac mass, and the surface derivative appearing in the spin-orbit potential, and is too rigid to reconcile them without additional field couplings. A qualitative improvement was seen by exploring the NJL-inspired scalar potential. Therefore, one path would be to consider a more systematic study within this approach.

A related limitation was also observed in the astrophysical sector, where RMF-CC tends to favor a larger slope of the symmetry energy, leading to tensions with astrophysical constraints when the model is calibrated to neutron-star observables~\cite{Pradhan2026}. This suggests that additional isovector flexibility may be required. A natural phenomenological extension would be to introduce cross-meson couplings, in particular a vector-isovector coupling, as explored in Ref.~\cite{Pradhan2026}, and to perform a joint calibration to finite-nucleus observables, NEPs, and astrophysical constraints. \\
Other natural extensions that could affect the spin-orbit potential include adding, at the Hartree level, a tensor isoscalar term, as in Ref.\cite{Mercier-spinorbit}, or, more microscopically, adding the Fock terms. As shown in Ref.~\cite{Chanfray_rho}, the tensor $\rho$ interaction plays an important role in the strength of the spin-orbit potential, and relativistic Hartree-Fock (RHF) studies for infinite matter, such as in Ref.~\cite{Chamseddine}, seem to point to a reduction of the Dirac and NR effective masses when Fock terms are included. A natural next step is therefore to include the Fock terms, and include a systematic refit of the pairing strength to some spectroscopic data. This can provide a valuable extension towards a more complete and accurate energy density functional for finite nuclei and nuclear matter.

{\bf Acknowledgments:}
The authors acknowledge the support of the project RELANSE ANR-23-CE31-0027-01 of the French National Research Agency (ANR), the CNRS-IN2P3 MAC2 masterproject, the European Union’s Horizon 2020 research and innovation program under grant agreement STRONG–2020-No824093.

\bibliographystyle{spphys}  
\bibliography{biblio}

@article{Bender:2003,
  title = {Self-consistent mean-field models for nuclear structure},
  author = {Bender, Michael and Heenen, Paul-Henri and Reinhard, Paul-Gerhard},
  journal = {Rev. Mod. Phys.},
  volume = {75},
  issue = {1},
  pages = {121--180},
  numpages = {0},
  year = {2003},
  month = {Jan},
  publisher = {American Physical Society},
  doi = {10.1103/RevModPhys.75.121},
  url = {https://link.aps.org/doi/10.1103/RevModPhys.75.121}
}

@article{Chamseddine,
   title={Relativistic Hartree–Fock chiral Lagrangians with confinement, nucleon finite size and short-range effects},
   volume={59},
   ISSN={1434-601X},
   url={http://dx.doi.org/10.1140/epja/s10050-023-01089-2},
   doi={10.1140/epja/s10050-023-01089-2},
   number={8},
   journal={The European Physical Journal A},
   publisher={Springer Science and Business Media LLC},
   author={Chamseddine, Mohamad and Margueron, Jérôme and Chanfray, Guy and Hansen, Hubert and Somasundaram, Rahul},
   year={2023},
   month=aug }

@article{Chamseddine_NJL,
  author = {Chamseddine, M. and Margueron, J. and Hansen, H. and Chanfray, G.},
  title   = {Hartree--Fock Lagrangians with a Nambu--Jona--Lasinio scalar potential},
  journal = {The European Physical Journal A},
  volume  = {60},
  pages   = {137},
  year    = {2024},
  doi     = {10.1140/epja/s10050-024-01358-8}
}

@article{RING1996,
  author  = {Ring, P.},
  title   = {Relativistic mean field theory in finite nuclei},
  journal = {Prog. Part. Nucl. Phys.},
  volume  = {37},
  pages   = {193--263},
  year    = {1996},
  doi     = {10.1016/0146-6410(96)00054-3}
}

@article{NIKSIC_code,
title = {DIRHB—A relativistic self-consistent mean-field framework for atomic nuclei},
journal = {Computer Physics Communications},
volume = {185},
number = {6},
pages = {1808-1821},
year = {2014},
issn = {0010-4655},
doi = {https://doi.org/10.1016/j.cpc.2014.02.027},
url = {https://www.sciencedirect.com/science/article/pii/S0010465514000836},
author = {T. Nikšić and N. Paar and D. Vretenar and P. Ring},
}

@article{binding_energies,
author = {Audi, Georges and Kondev, Filip and Wang, Meng and Huang, Wenjia and Naimi, S.},
year = {2017},
month = {03},
pages = {030001},
title = {The NUBASE2016 evaluation of nuclear properties},
volume = {41},
journal = {Chinese Physics C},
doi = {10.1088/1674-1137/41/3/030001}
}

@article{charge_radii,
author = {Angeli, István and Marinova, Krassimira},
year = {2013},
month = {01},
pages = {69-95},
title = {Table of experimental nuclear ground state charge radii: An update},
volume = {99},
journal = {Atomic Data and Nuclear Data Tables},
doi = {10.1016/j.adt.2011.12.006}
}

@article{Odilon2023,
  title = {Low-energy nuclear physics and global neutron star properties},
  author = {Carlson, Brett V. and Dutra, Mariana and Louren\ifmmode \mbox{\c{c}}\else \c{c}\fi{}o, Odilon and Margueron, J\'er\^ome},
  journal = {Phys. Rev. C},
  volume = {107},
  issue = {3},
  pages = {035805},
  numpages = {24},
  year = {2023},
  month = {Mar},
  publisher = {American Physical Society},
  doi = {10.1103/PhysRevC.107.035805},
  url = {https://link.aps.org/doi/10.1103/PhysRevC.107.035805}
}

@article{Meng2006,
  author  = {Meng, J. and Toki, H. and Zhou, S.-G. and Zhang, S.-Q. and Long, W.-H. and Geng, L.-S.},
  title   = {Relativistic Continuum {Hartree--Bogoliubov} Theory for Ground-State Properties of Exotic Nuclei},
  journal = {Progress in Particle and Nuclear Physics},
  volume  = {57},
  number  = {2},
  pages   = {470--563},
  year    = {2006},
  doi     = {10.1016/j.ppnp.2005.06.001}
}

@article{VRETENAR2005,
title = {Relativistic Hartree–Bogoliubov theory: static and dynamic aspects of exotic nuclear structure},
journal = {Physics Reports},
volume = {409},
number = {3},
pages = {101-259},
year = {2005},
issn = {0370-1573},
doi = {https://doi.org/10.1016/j.physrep.2004.10.001},
author = {D. Vretenar and A.V. Afanasjev and G.A. Lalazissis and P. Ring},
}

@article{Dobaczewski:2014,
doi = {10.1088/0954-3899/41/7/074001},
url = {https://dx.doi.org/10.1088/0954-3899/41/7/074001},
year = {2014},
month = {may},
publisher = {IOP Publishing},
volume = {41},
number = {7},
pages = {074001},
author = {J Dobaczewski and W Nazarewicz and P-G Reinhard},
title = {Error estimates of theoretical models: a guide},
journal = {Journal of Physics G: Nuclear and Particle Physics}
}

@article{Ebrand-spinorbit,
  title = {Spin-orbit interaction in relativistic nuclear structure models},
  author = {Ebran, J.-P. and Mutschler, A. and Khan, E. and Vretenar, D.},
  journal = {Phys. Rev. C},
  volume = {94},
  issue = {2},
  pages = {024304},
  numpages = {7},
  year = {2016},
  month = {Aug},
  publisher = {American Physical Society},
  doi = {10.1103/PhysRevC.94.024304},
  url = {https://link.aps.org/doi/10.1103/PhysRevC.94.024304}
}

@article{Mercier-spinorbit,
  title = {Covariant energy density functionals with and without tensor couplings at the Hartree-Bogoliubov level},
  author = {Mercier, F. and Ebran, J.-P. and Khan, E.},
  journal = {Phys. Rev. C},
  volume = {107},
  issue = {3},
  pages = {034309},
  numpages = {14},
  year = {2023},
  month = {Mar},
  publisher = {American Physical Society},
  doi = {10.1103/PhysRevC.107.034309},
  url = {https://link.aps.org/doi/10.1103/PhysRevC.107.034309}
}

@article{Ray1979,
  title = {Neutron densities and the single particle structure of several even-even nuclei from $^{40}\mathrm{Ca}$ to $^{208}\mathrm{Pb}$},
  author = {Ray, L. and Hodgson, P. E.},
  journal = {Phys. Rev. C},
  volume = {20},
  issue = {6},
  pages = {2403--2417},
  numpages = {0},
  year = {1979},
  month = {Dec},
  publisher = {American Physical Society},
  doi = {10.1103/PhysRevC.20.2403},
  url = {https://link.aps.org/doi/10.1103/PhysRevC.20.2403}
}

@article{MONTANARI2011,
title = {Probing the nature of particle–core couplings in 49Ca with γ spectroscopy and heavy-ion transfer reactions},
journal = {Physics Letters B},
volume = {697},
number = {4},
pages = {288-293},
year = {2011},
issn = {0370-2693},
doi = {https://doi.org/10.1016/j.physletb.2011.01.046},
url = {https://www.sciencedirect.com/science/article/pii/S0370269311000839},
author = {D. Montanari and S. Leoni and D. Mengoni and G. Benzoni and N. Blasi and G. Bocchi and P.F. Bortignon and A. Bracco and F. Camera and G. Colò and A. Corsi and F.C.L. Crespi and B. Million and R. Nicolini and O. Wieland and J.J. Valiente-Dobon and L. Corradi and G. {de Angelis} and F. {Della Vedova} and E. Fioretto and A. Gadea and D.R. Napoli and R. Orlandi and F. Recchia and E. Sahin and R. Silvestri and A.M. Stefanini and R.P. Singh and S. Szilner and D. Bazzacco and E. Farnea and R. Menegazzo and A. Gottardo and S.M. Lenzi and S. Lunardi and G. Montagnoli and F. Scarlassara and C. Ur and G. {Lo Bianco} and A. Zucchiatti and M. Kmiecik and A. Maj and W. Meczynski and A. Dewald and Th. Pissulla and G. Pollarolo},
}

@misc{chanfray2026,
      title={Mechanical properties of the nucleon in the chiral confining model. II -- in-medium evolution of the nucleon properties}, 
      author={Guy Chanfray and Hubert Hansen and Bikram Keshari Pradhan},
      year={2026},
      eprint={2606.03633},
      archivePrefix={arXiv},
      primaryClass={nucl-th},
      url={https://arxiv.org/abs/2606.03633}, 
}

@Article{Chanfray_CCM,
AUTHOR = {Chanfray, Guy and Ericson, Magda and Hansen, Hubert and Margueron, Jérôme and Martini, Marco},
TITLE = {From QCD Phenomenology to Nuclear Physics Phenomenology: The Chiral Confining Model},
JOURNAL = {Symmetry},
VOLUME = {17},
YEAR = {2025},
NUMBER = {2},
ARTICLE-NUMBER = {313},
URL = {https://www.mdpi.com/2073-8994/17/2/313},
ISSN = {2073-8994},
DOI = {10.3390/sym17020313}
}

@article{Margueron2019,
author = {Margueron, J. and Gulminelli, F.},
year = {2019},
month = {02},
pages = {},
title = {Effect of high-order empirical parameters on the nuclear equation of state},
volume = {99},
journal = {Physical Review C},
doi = {10.1103/PhysRevC.99.025806}
}

@article{Chanfray_Cchi,
  TITLE = {{Constraints on the in-medium nuclear interaction from chiral symmetry and Lattice-QCD}},
  AUTHOR = {Chanfray, G and Hansen, H and Margueron, J},
  URL = {https://hal.science/hal-04074564},
  JOURNAL = {{Eur.Phys.J.A}},
  VOLUME = {59},
  NUMBER = {11},
  PAGES = {264},
  YEAR = {2023},
  DOI = {10.1140/epja/s10050-023-01179-1},
  HAL_ID = {hal-04074564},
  HAL_VERSION = {v1},
}

@article{Chanfray-Schuck,
    author = "Chanfray, Guy",
    title = "{Scalar field, nucleon structure and relativistic chiral theory for nuclear matter}",
    eprint = "2310.19532",
    archivePrefix = "arXiv",
    primaryClass = "nucl-th",
    doi = "10.1140/epja/s10050-023-01221-2",
    journal = "Eur. Phys. J. A",
    volume = "60",
    number = "1",
    pages = "7",
    year = "2024"
}

@Article{chanfray-universe,
AUTHOR = {Chanfray, Guy and Ericson, Magda and Martini, Marco},
TITLE = {The Interrelated Roles of Correlations in the Nuclear Equation of State and in Response Functions: Application to a Chiral Confining Theory},
JOURNAL = {Universe},
VOLUME = {9},
YEAR = {2023},
NUMBER = {7},
ARTICLE-NUMBER = {316},
URL = {https://www.mdpi.com/2218-1997/9/7/316},
ISSN = {2218-1997},
DOI = {10.3390/universe9070316}
}

@PREAMBLE{
 "\providecommand{\noopsort}[1]{}" 
 # "\providecommand{\singleletter}[1]{#1}%" 
}

@article{Rahul2022,
    author = "Somasundaram, Rahul and Margueron, J\'er\^ome and Chanfray, Guy and Hansen, Hubert",
    title = "{Comparison of different relativistic models applied to dense nuclear matter}",
    eprint = "2109.05374",
    archivePrefix = "arXiv",
    primaryClass = "nucl-th",
    doi = "10.1140/epja/s10050-022-00733-7",
    journal = "Eur. Phys. J. A",
    volume = "58",
    number = "5",
    pages = "84",
    year = "2022"
}

@article{Hebeler2013,
    author = "Hebeler, K. and Lattimer, J. M. and Pethick, C. J. and Schwenk, A.",
    title = "{Equation of state and neutron star properties constrained by nuclear physics and observation}",
    eprint = "1303.4662",
    archivePrefix = "arXiv",
    primaryClass = "astro-ph.SR",
    doi = "10.1088/0004-637X/773/1/11",
    journal = "Astrophys. J.",
    volume = "773",
    pages = "11",
    year = "2013"
}

@Article{Nambu:1961fr,
     author    = "Nambu, Yoichiro and Jona-Lasinio, G.",
     title     = "{Dynamical model of elementary particles based on an
                  analogy with  superconductivity. II}",
     journal   = "Phys. Rev.",
     volume    = "124",
     year      = "1961",
     pages     = "246-254",
     doi       = "10.1103/PhysRev.124.246",
     SLACcitation  = "%%CITATION = PHRVA,124,246;%%"                                                                                                          
}

@article{Chanfray2001,
    author = "Chanfray, G. and Ericson, Magda and Guichon, Pierre A. M.",
    title = "{Chiral symmetry and quantum hadrodynamics}",
    eprint = "nucl-th/0012013",
    archivePrefix = "arXiv",
    reportNumber = "LYCEN-2000-130, CERN-TH-2000-346, DAPNIA-SPHN-00-74",
    doi = "10.1103/PhysRevC.63.055202",
    journal = "Phys. Rev. C",
    volume = "63",
    pages = "055202",
    year = "2001"
}

@article{Chanfray2005,
    author = "Chanfray, G. and Ericson, M.",
    title = "{QCD susceptibilities and nuclear-matter saturation in a relativistic chiral theory}",
    doi = "10.1140/epja/i2005-10074-6",
    journal = "Eur. Phys. J. A",
    volume = "25",
    pages = "151--157",
    year = "2005"
}

@article{Weinberg,
    author = "Weinberg, Steven",
    title = "{Nuclear forces from chiral Lagrangians}",
    reportNumber = "UTTG-31-90",
    doi = "10.1016/0370-2693(90)90938-3",
    journal = "Phys. Lett. B",
    volume = "251",
    pages = "288--292",
    year = "1990"
}

@article{Chanfray2011,
    author = "Chanfray, G. and Ericson, M.",
    title = "{Scalar field in nuclear matter: the roles of spontaneous chiral symmetry breaking and nucleon structure}",
    eprint = "1011.4280",
    archivePrefix = "arXiv",
    primaryClass = "nucl-th",
    reportNumber = "LYCEN-2010-20",
    doi = "10.1103/PhysRevC.83.015204",
    journal = "Phys. Rev. C",
    volume = "83",
    pages = "015204",
    year = "2011"
}

@article{Chanfray2006,
    author = "Chanfray, G. and Davesne, D. and Ericson, M. and Martini, M.",
    title = "{Two-pion production processes, chiral symmetry and N N interaction in the medium}",
    eprint = "nucl-th/0406003",
    archivePrefix = "arXiv",
    reportNumber = "LYCEN-2004-08",
    doi = "10.1140/epja/i2005-10245-5",
    journal = "Eur. Phys. J. A",
    volume = "27",
    pages = "191--198",
    year = "2006"
}

@article{Birse94,
    author = "Birse, Michael C.",
    title = "{What does a change in the quark condensate say about restoration of chiral symmetry in matter?}",
    eprint = "hep-ph/9602266",
    archivePrefix = "arXiv",
    reportNumber = "MC-TH-96-09",
    doi = "10.1103/PhysRevC.53.R2048",
    journal = "Phys. Rev. C",
    volume = "53",
    pages = "R2048--R2051",
    year = "1996"
}

@article{Guichon1988,
    author = "Guichon, Pierre A. M.",
    title = "{A Possible Quark Mechanism for the Saturation of Nuclear Matter}",
    reportNumber = "LYCEN-8762",
    doi = "10.1016/0370-2693(88)90762-9",
    journal = "Phys. Lett. B",
    volume = "200",
    pages = "235--240",
    year = "1988"
}

@article{Guichon2004,
    author = "Guichon, Pierre A. M. and Thomas, Anthony William",
    title = "{Quark structure and nuclear effective forces}",
    eprint = "nucl-th/0402064",
    archivePrefix = "arXiv",
    reportNumber = "DAPNIA-04-37, ADP-04-03-T582",
    doi = "10.1103/PhysRevLett.93.132502",
    journal = "Phys. Rev. Lett.",
    volume = "93",
    pages = "132502",
    year = "2004"
}

@article{vanDalen2005,
  title = {Momentum, density, and isospin dependence of symmetric and asymmetric nuclear matter properties},
  author = {Dalen, E. N. E. van and Fuchs, C. and Faessler, Amand},
  journal = {Phys. Rev. C},
  volume = {72},
  issue = {6},
  pages = {065803},
  numpages = {9},
  year = {2005},
  month = {Dec},
  publisher = {American Physical Society},
  doi = {10.1103/PhysRevC.72.065803},
}

@article{Lalazissis2005,
    author = "Lalazissis, G. A. and Niksic, T. and Vretenar, D. and Ring, P.",
    title = "{New relativistic mean-field interaction with density-dependent meson-nucleon couplings}",
    doi = "10.1103/PhysRevC.71.024312",
    journal = "Phys. Rev. C",
    volume = "71",
    pages = "024312",
    year = "2005"
}

@article{Long2007,
    author = "Long, Wen Hui and Sagawa, Hiroyuki and Van Giai, Nguyen and Meng, Jie",
    title = "{Shell Structure and rho-Tensor Correlations in Density-Dependent Relativistic Hartree-Fock theory}",
    eprint = "0706.3497",
    archivePrefix = "arXiv",
    primaryClass = "nucl-th",
    doi = "10.1103/PhysRevC.76.034314",
    journal = "Phys. Rev. C",
    volume = "76",
    pages = "034314",
    year = "2007"
}

@article{Chanfray2007,
    author = "Chanfray, G. and Ericson, M.",
    title = "{QCD susceptibilities and nuclear matter saturation in a chiral theory: inclusion of pion loops}",
    eprint = "nucl-th/0611042",
    archivePrefix = "arXiv",
    reportNumber = "LYCEN-2006-20",
    doi = "10.1103/PhysRevC.75.015206",
    journal = "Phys. Rev. C",
    volume = "75",
    pages = "015206",
    year = "2007"
}

@article{Massot2008,
    author = "Massot, E. and Chanfray, G.",
    title = "{Relativistic Chiral Hartree-Fock description of nuclear matter with constraints from nucleon structure and confinement}",
    eprint = "0803.1719",
    archivePrefix = "arXiv",
    primaryClass = "nucl-th",
    reportNumber = "LYCEN-2008-02",
    doi = "10.1103/PhysRevC.78.015204",
    journal = "Phys. Rev. C",
    volume = "78",
    pages = "015204",
    year = "2008"
}

@article{Massot2009,
    author = "Massot, E. and Chanfray, G.",
    title = "{Relativistic calculation of the pion loop correlation energy in nuclear matter in a theory including confinement}",
    eprint = "0905.0605",
    archivePrefix = "arXiv",
    primaryClass = "nucl-th",
    reportNumber = "LYCEN-2009-05",
    doi = "10.1103/PhysRevC.80.015202",
    journal = "Phys. Rev. C",
    volume = "80",
    pages = "015202",
    year = "2009"
}

@article{Margueron2018,
    author = "Margueron, J\'er\^ome and Hoffmann Casali, Rudiney and Gulminelli, Francesca",
    title = "{Equation of state for dense nucleonic matter from metamodeling. I. Foundational aspects}",
    eprint = "1708.06894",
    archivePrefix = "arXiv",
    primaryClass = "nucl-th",
    reportNumber = "INT-PUB-17-029",
    doi = "10.1103/PhysRevC.97.025805",
    journal = "Phys. Rev. C",
    volume = "97",
    number = "2",
    pages = "025805",
    year = "2018"
}

@article{Lalazissis1996,
    author = "Lalazissis, G. A. and Konig, J. and Ring, P.",
    title = "{A New parametrization for the Lagrangian density of relativistic mean field theory}",
    eprint = "nucl-th/9607039",
    archivePrefix = "arXiv",
    doi = "10.1103/PhysRevC.55.540",
    journal = "Phys. Rev. C",
    volume = "55",
    pages = "540--543",
    year = "1997"
}

@article{BERGER1991,
title = {Time-dependent quantum collective dynamics applied to nuclear fission},
journal = {Computer Physics Communications},
volume = {63},
number = {1},
pages = {365-374},
year = {1991},
issn = {0010-4655},
doi = {https://doi.org/10.1016/0010-4655(91)90263-K},
url = {https://www.sciencedirect.com/science/article/pii/001046559190263K},
author = {J.F. Berger and M. Girod and D. Gogny},
}

@article{charge_density,
title = {Nuclear charge-density-distribution parameters from elastic electron scattering},
journal = {Atomic Data and Nuclear Data Tables},
volume = {36},
number = {3},
pages = {495-536},
year = {1987},
issn = {0092-640X},
doi = {https://doi.org/10.1016/0092-640X(87)90013-1},
url = {https://www.sciencedirect.com/science/article/pii/0092640X87900131},
author = {H. {De Vries} and C.W. {De Jager} and C. {De Vries}},
}

@article{Yao2012,
    author = "Yao, Jiang-Ming and Baroni, Simone and Bender, Michael and Heenen, Paul-Henri",
    title = "{Beyond-mean-field study of the possible 'bubble' structure of 34Si}",
    eprint = "1205.2262",
    archivePrefix = "arXiv",
    primaryClass = "nucl-th",
    doi = "10.1103/PhysRevC.86.014310",
    journal = "Phys. Rev. C",
    volume = "86",
    pages = "014310",
    year = "2012"
}

@article{Chanfray_rho,
  title = {Contribution of the $\ensuremath{\rho}$ meson and quark substructure to the nuclear spin-orbit potential},
  author = {Chanfray, Guy and Margueron, J\'er\^ome},
  journal = {Phys. Rev. C},
  volume = {102},
  issue = {2},
  pages = {024331},
  numpages = {9},
  year = {2020},
  month = {Aug},
  publisher = {American Physical Society},
  doi = {10.1103/PhysRevC.102.024331},
  url = {https://link.aps.org/doi/10.1103/PhysRevC.102.024331}
}

@article{PDM_Kim,
    author = "Mun, Myeong-Hwan and Shin, Ik Jae and Paeng, Won-Gi and Harada, Masayasu and Kim, Youngman",
    title = "{Nuclear structure in parity doublet model}",
    eprint = "1805.03402",
    archivePrefix = "arXiv",
    primaryClass = "nucl-th",
    doi = "10.1140/epja/s10050-023-01064-x",
    journal = "Eur. Phys. J. A",
    volume = "59",
    number = "7",
    pages = "149",
    year = "2023"
}

@article{Ma2004,
    author = "Ma, Zhong-yu and Rong, Jian and Chen, Bao-Qiu and Zhu, Zhi-Yuan and Song, Hong-Qiu",
    title = "{Isospin dependence of nucleon effective mass in Dirac Brueckner-Hartree-Fock approach}",
    eprint = "nucl-th/0412030",
    archivePrefix = "arXiv",
    doi = "10.1016/j.physletb.2004.11.004",
    journal = "Phys. Lett. B",
    volume = "604",
    pages = "170--174",
    year = "2004"
}

@article{Jaminon1989,
    author = "Jaminon, M. and Mahaux, C.",
    title = "{Effective Masses in Relativistic Approaches to the Nucleon Nucleus Mean Field}",
    doi = "10.1103/PhysRevC.40.354",
    journal = "Phys. Rev. C",
    volume = "40",
    pages = "354--367",
    year = "1989"
}

@ARTICLE{Pradhan2026,
       author = {{Keshari Pradhan}, Bikram and {Chamseddine}, Mohamad and {Margueron}, J{\'e}r{\^o}me and {Hansen}, Hubert and {Chanfray}, Guy and {Ebran}, Jean-Paul and {Khan}, Elias},
        title = "{Relativistic Mean Field Approach with Chiral Symmetry Breaking and Quark Confinement in the light of Astrophysical Observations}",
      journal = {arXiv e-prints},
         year = 2026,
        month = jul,
          eid = {arXiv:2607.08412},
        pages = {arXiv:2607.08412},
          doi = {10.48550/arXiv.2607.08412},
archivePrefix = {arXiv},
       eprint = {2607.08412},
 primaryClass = {nucl-th},
       adsurl = {https://ui.adsabs.harvard.edu/abs/2026arXiv260708412K}
}

\end{document}